\documentclass[10pt,journal]{IEEEtran}
\usepackage{multirow}
\usepackage{tablefootnote}
\usepackage{graphicx}
\usepackage{amsfonts}
\usepackage{amsmath,amssymb}
\usepackage{units}
\usepackage{booktabs}
\usepackage{comment}
\usepackage[binary-units=true]{siunitx}
\usepackage{amsfonts}
\usepackage{multirow}
\usepackage{amsmath,amssymb}
\usepackage{amsmath}
\usepackage{ulem}
\usepackage{cite}
\usepackage{units}
\usepackage{algorithm}
\usepackage{graphicx}
\usepackage{color}
\usepackage{booktabs}
\usepackage{algpseudocode}
\usepackage{enumitem,kantlipsum}
\usepackage{algpseudocode}
\usepackage{hyperref}
\algdef{SE}[DOWHILE]{Do}{doWhile}{\algorithmicdo}[1]{\algorithmicwhile\ #1}%
\usepackage{xcolor}

\usepackage{hyperref}

\renewcommand{\baselinestretch}{0.99}

\usepackage{tikz}
\newcommand*\circled[1]{\tikz[baseline=(char.base)]{
            \node[shape=circle,fill,inner sep=1pt] (char) {\textcolor{white}{#1}};}}
            
\newcommand{\norm}[1]{\left\lVert#1\right\rVert}

\newcommand{\RSNR}{\rm{RSNR }}

\begin{document}
\title{Exploring Scalable, Distributed Real-Time Anomaly Detection for Bridge Health Monitoring}
%\title{Anomaly detection for Bridge health monitoring, edge vs cloud computing}
%
%
\author{Amirhossein~Moallemi,
        Alessio~Burrello,~\IEEEmembership{Graduate Student Member,~IEEE,} \\Davide~Brunelli,~\IEEEmembership{Senior Member,~IEEE,}
        and~Luca~Benini,~\IEEEmembership{Fellow,~IEEE}% <-this % stops a space
\IEEEcompsocitemizethanks{\IEEEcompsocthanksitem A. Burrello, A. Moallemi, D. Brunelli, and L. Benini are with the Department of Electrical, Electronic and Information Engineering, University of Bologna, 40136 Bologna, Italy.\protect\\
E-mail: alessio.burrello@unibo.it, amirhossein.moallem2@unibo.it
\IEEEcompsocthanksitem D. Brunelli is also with the Department of Industrial Engineering, University of Trento, 38123 Trento, Italy.\protect\\
E-mail: davide.brunelli@unitn.it
\IEEEcompsocthanksitem L. Benini is also with the Department of Information Technology and Electrical Engineering at the ETH Zurich, 8092 Zurich, Switzerland.\protect\\
E-mail: lbenini@iis.ee.ethz.ch} % <-this % stops a space
\thanks{This work was supported by the Italian Ministry for University and Research (MUR) under the program “Dipartimenti di Eccellenza (2018-2022)” and partially supported by the EU H2020-ECSEL project Arrowhead Tools (g.a. 826452). Moreover the authors would like to thank Sacertis S.r.l. for the dataset.}
\thanks{Manuscript received October, 2021.}
\thanks{Copyright (c) 20xx IEEE. Personal use of this material is permitted. However, permission to use this material for any other purposes must be obtained from the IEEE by sending a request to \url{pubs-permissions@ieee.org}}
\thanks{Link to the github repository of this paper: \url{https://github.com/MiirHo3eIN/Real_Time_BHM}}
\thanks{This article has been accepted for publication in the IEEE Internet of Things Journal.  doi: 10.1109/JIOT.2022.3157532}
}

%\markboth{IEEE Internet of Things Journal,~Vol.~xx, No.~xx, October~2021}%
%{Shell \MakeLowercase{\textit{et al.}}: Bare Demo of IEEEtran.cls for Computer Society Journals}

\IEEEtitleabstractindextext{%
\begin{abstract}
Modern real-time Structural Health Monitoring systems can generate a considerable amount of information that must be processed and evaluated for detecting early anomalies and generating prompt warnings and alarms about the civil infrastructure conditions. 
The current cloud-based solutions cannot scale if the raw data has to be collected from thousands of buildings.
This paper presents a full-stack deployment of an efficient and scalable anomaly detection pipeline for SHM systems which does not require sending raw data to the cloud but relies on edge computation. First, we benchmark three algorithmic approaches of anomaly detection, i.e., Principal Component Analysis (PCA), Fully-Connected AutoEncoder (FC-AE), and Convolutional AutoEncoder (C-AE).
Then, we deploy them on an edge-sensor, the STM32L4, with limited computing capabilities. Our approach decreases network traffic by $\approx8\cdot10^5\times$
, from 780KB/hour to less than 10 Bytes/hour for a single installation and minimize network and cloud resource utilization, enabling the scaling of the monitoring infrastructure.  
A real-life case study, a highway bridge in Italy, demonstrates that combining near-sensor computation of anomaly detection algorithms, smart pre-processing, and low-power wide-area network protocols (LPWAN) we can greatly reduce data communication and cloud computing costs, while anomaly detection accuracy is not adversely affected.
\end{abstract}

% Note that keywords are not normally used for peer review papers.
}
\maketitle
\IEEEdisplaynontitleabstractindextext

\section{Introduction}
\label{sec:introduction} 
The constant growth of large-scale civil infrastructures built worldwide in recent years~\cite{ref1-100, book1-1} is sustained by increasing investments of the top economies in the world for renewing urban areas and roads.
Furthermore, civil infrastructures are becoming increasingly complex. For instance, nowadays, new technologies permit the construction of long-span viaducts, long-extension undersea tunnels, and large skyscraper districts. 

On the other hand, although recent structural design achievements assure robustness in these advanced buildings, vulnerability to diverse threats such as extreme weather (e.g., wind, rain), massive vehicular load, and earthquakes remain a major concern~\cite {Ref1-3,Ref1-1}.
The recent \textit{Polcevera viaduct's collapse} in Genoa, Italy, with over 40 human lives lost,  demonstrated that the periodic or sporadic human-assisted assessment of the structures is not enough.
Therefore, a change of paradigm towards the continuous observation of structural integrity and automatic anomaly detection is becoming a key requirement for civil infrastructure maintenance~\cite{Ref1-6, Ref1-7}.    
\begin{figure}
  \centering
\includegraphics[width=0.9\columnwidth]{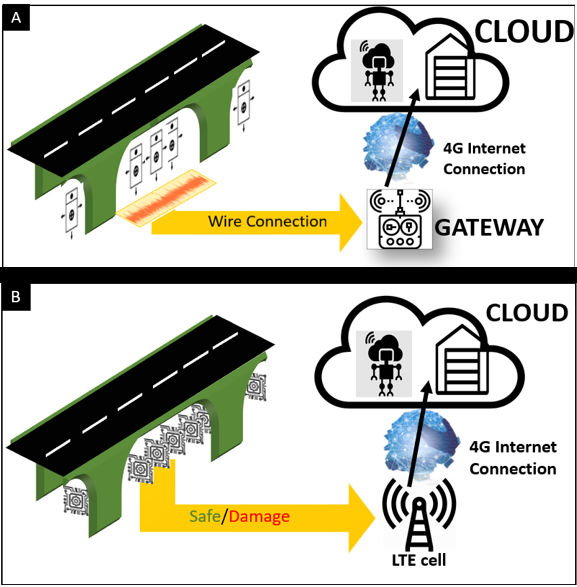}
  \caption{IoT based SHM systems. In Panel A, the raw signal is gathered from the sensors and sent to the cloud through a gateway to analyse the condition of the structure. In Panel B, the safe/damage condition is directly computed on the node.}
\label{fig:trd_vs_modern}
 \vspace{-0.5cm}
\end{figure}

As a consequence, the new field of automated Structural Health Monitoring (SHM), which tracks the real-time online state of structures using dense sensor networks, is gaining prominence~\cite{Ref1-9, chang2019real}.
Combined with the advancements in the Internet of Things (IoT) \cite{arslan2020advancements}, SHM enables monitoring large structures at smaller costs than by deploying human crews~\cite{ref1-14-new}.

A modern SHM system is composed of a series of sensor-nodes, which capture sensor data, e.g. the vibration of a viaduct, one or more data collection and processing gateways, and centralized processing and storage resources in the cloud~\cite{ref1-17-new}.
Fig. \ref{fig:trd_vs_modern}A depicts a SHM installation on a viaduct.
The amount of information and data gathered by new generation SHM systems is exponentially growing, moving from a few measurements every hour from few sensors to continuous high-frequency data streams from dense sensors networks~\cite{ref1-16, ref1-27}. 
Therefore, data communication and storage capabilities in the cloud have become major concerns in modern SHM systems. 

This paper focuses on two key challenges: automating anomaly detection and doing so with a scalable approach that does not require communication, processing and storing raw sensor data in the cloud. 
For anomaly detection, different techniques have been proposed, ranging from simple regressive models~\cite{ref2-18} to deep neural networks~\cite{ref2-1-new}.
However, they are usually tested on simulated data, not taking into account real-condition perturbations such as wind or climate fluctuations~\cite{ref1-22, ref1-23}.
Besides that, these techniques are typically deployed on cloud servers. Hence they imply comprehensive data collection from the sensor network.

In this work, we address both challenges by proposing a new pipeline for viaduct monitoring.
We analyze a real highway viaduct in Italy, which underwent a pre-scheduled maintenance intervention. The viaduct has been monitored before and after the intervention, and acceleration data from five sensors has been collected. We use this real-life example of an exogenous event changing the structural properties of the bridge as a proxy for an abrupt unforeseen event such as an earthquake. Specifically, we give the following contributions: 
\begin{itemize}
    \item We compare data-driven and model-driven unsupervised anomaly detection approaches to monitor the behaviour of the viaduct: namely a Principal Component Analysis (PCA) model and two autoencoders. On our dataset, the PCA shows the best performance with $98.8\%$ accuracy in detecting the structural changes after the interventions. Further, we assessed in depth our anomaly detection approach robustness by synthetically generating new anomalies from the original real-life one.
    \item We find that the best accuracy (100\%) on anomaly detection is obtained by computing PCA-based reconstruction error on 5 seconds time windows and averaging errors over 4 hours, i.e., 2880 windows. Noteworthy, this result comes at the cost of a non-negligible delay in detecting the damage (4 hours). This trade-off between accuracy and delay can be tuned by the user by simply changing the length of the averaging window.
    \item We study the impact of energy-threshold-based pre-filtering of the raw accelerometer data, showing that this preprocessing block is essential to achieve high anomaly detection accuracy while decreasing computational effort by 17\%. 
    \item We compare the performance of the algorithms when fed with time-domain (raw) or frequency-domain (processed with FFT) data. We observe a drop of $ 21.58\%$ and $ 12.73\%$ of anomaly detection accuracy, respectively, for both the PCA and Fully Connected (FC) autoencoder, when using frequency domain data. 
    \item We present the implementation of our anomaly detection pipeline on a low-power microcontroller for online inference with $\approx74$ uJ energy consumption for each inference. We show the trade-off between accuracy and power consumption by tuning the hyperparameters of our best performing anomaly detector, the PCA. We show that by increasing the Compression Factor (CF) of the PCA (i.e., by reducing the number of principal components utilized during the reconstruction) from 16 to 24, we still achieve $92.97\%$ accuracy in detecting structural changes (i.e., distinguishing anomalies from normal samples) while consuming only $39$ uJ per inference.
    \item We demonstrate our distributed approach using a node equipped with a Narrowband IoT communication unit \cite{ref1-28}, which exploits anomaly detection at the edge. We show that by performing computation on the edge and communicating only post-processed data (i.e., detected anomalies), the energy consumption of a node can be reduced by $5.0\times$.
\end{itemize}
The paper is organized as follows. 
Sec. \ref{sec:relatedWork} describes the related work. 
Sec. \ref{sec:bakground} introduces the viaduct structure, the SHM framework installed, and the background on the anomaly detection methods proposed in this work.
Sec. \ref{sec:scenarios} describes our proposed pipeline, with different analyzed design and deployment choices.
Sec. \ref{sec:software_results} and Sec. \ref{sec:deployment_results} provide results of both the anomaly detector accuracy and the deployment-related metrics, i.e., memory footprint, energy consumption, and network data communication.
Sec. \ref{sec:Conclusion} concludes the paper. 

\section{Related Work}
\label{sec:relatedWork}
\begin{table*}[]
\centering
\caption{Structural Health Monitoring studies over the last years. Performance results refer to the distinction of damaged from non-damaged data samples. Performance is in terms of Accuracy unless it is mentioned. Abbreviations: FP: False positives, FN: False Negative.}
\label{tab:related_work}
\renewcommand{\arraystretch}{1.3}
\begin{tabular}{l|llll|ll|l}
                             & Structure                                                                      & Sensors                                   & Data Type                                                               & Detection Model           & Train Device  & Test Device   & Performance                                                    \\ \hline

\multicolumn{8}{l}{\textbf{Statistical Data Modelling}}    \\\hline
Ling et al. \cite{ref2-41}    & \begin{tabular}[c]{@{}l@{}}Simulated Steel\\ Frame Structure\end{tabular}                                              & 120                                            & Acceleration                                                            & \begin{tabular}[c]{@{}l@{}}Autoregressive \\model\end{tabular}      & Remote Server & Remote Server~\cite{OpenSees}  & 1 FP, 1 FN                                                     \\\hline
Santos et al. \cite{ref2-42} & \begin{tabular}[c]{@{}l@{}}Simulated Five\\ Bay Structure \end{tabular}                                                  & 4$\times$29                                    & Acceleration                                                            & \begin{tabular}[c]{@{}l@{}}FFT + \\Peak Detection\end{tabular}         & N.A.          & MICAz \cite{micaz}       & 0 FN, 2 FP                                                     \\\hline

Verma et al. \cite{ref2-46}  & \begin{tabular}[c]{@{}l@{}}Simulated Steel Beam\\ Bridge\end{tabular}          & \begin{tabular}[c]{@{}l@{}}2\\ 14\end{tabular} & Acceleration                                                            & \begin{tabular}[c]{@{}l@{}}Features +\\ Gaussian Model\end{tabular} & N.A. & N.A.   & \begin{tabular}[c]{@{}l@{}}85-96\% \\ 96.3-100\% \end{tabular}           \\ \hline

\multicolumn{8}{l}{\textbf{Deep Neural Networks}}    \\\hline
Acvi et al. \cite{ref2-44}   & BM benchmark \cite{dyke2003experimental}                                       & 12                                             & Acceleration                                                            & 1D-CNN                    & Intel core-i7~\cite{Intel} & Intel core-i7~\cite{Intel}  & N.A.                                                           \\\hline
%Wu et al. \cite{ref2-45}     & \begin{tabular}[c]{@{}l@{}}Nuclear Power Plant\\ Metallic Surface\end{tabular} & 1                                              & \begin{tabular}[c]{@{}l@{}}Crack Images\\ Corrosion Images\end{tabular} & VGG/ResNet                & Intel Xeon~\cite{Intel_xenon} & Jetson TX2~\cite{nvidia}   & \begin{tabular}[c]{@{}l@{}}94.6\%-98.5\%\\ 82.8\%\end{tabular} \\\hline

\multicolumn{8}{l}{\textbf{Data reduction}}    \\\hline
Liu et al. \cite{ref2-29}                         & \begin{tabular}[c]{@{}l@{}}Simulated Lab-Scale\\  Bridge\end{tabular}          & 5                                              & Acceleration                                                            & Autoencoder                                    & N.A. & N.A.                      & N.A.                                                           \\ \hline
Nie et al. \cite{ref2-31}                         & \begin{tabular}[c]{@{}l@{}}Simulated Lab-Scale\\  Bridge\end{tabular}          & \begin{tabular}[c]{@{}l@{}}9\\ 24 \end{tabular}                                                & Acceleration                                                            & FMPCA                                    & Laptop & Remote Server                      & \begin{tabular}[c]{@{}l@{}}100\%\\ 100\% \end{tabular}    \\ \hline
\hline
Our Work                     & Real Viaduct                                                                   & 1                                              & Acceleration                                                            & AE / PCA         & STM32L476   & STM32L476   & 98.8\% \\\hline                                                       
\end{tabular}
\end{table*}
Structural Health Monitoring systems have become widespread in the last decade. They are usually based on networks of sensors to monitor the vibration of the structure under test~\cite{ref2-2-new}.
A key expected function of a modern SHM system is the automated detection of structural anomalies.
To solve this problem, we can distinguish between three main classes of approaches: i) statistical data modeling, ii) machine learning, and iii) data reduction approaches.

\subsubsection{Statistical data modeling}
The first approaches in continuous SHM systems were based on modeling data distribution and extracting abnormal patterns.
Ling et al.~\cite{ref2-41} exploit auto-regressive (AR) and auto-regressive with extra input (ARX) models to detect anomalies on a simulated steel frame structure to localized damage pattern recognition problems in SHM. The authors compute a set of statistical features on a cluster of nodes where sensors communicate via Random Gossip protocol to detect and localize the damage, implying that an individual node cannot detect damages to the structure. Although they report at most 1 False Negative (FN) and 1 False Positive (FP) detections, they performed experiments with four laboratory computed datasets.
Similarly, auto-regressive models are employed in e.g.,~\cite{ref2-51, ref2-52, ref2-53} to extract features from raw vibration data. 
One of the most recent works is Entezami, et al.~\cite{ref2-18} that propose an anomaly detection framework exploiting the recorded raw vibrations dataset of the Tianjin Yonghe cable-stayed bridge in China. First, an auto-regressive moving average (ARMA) extracts features to reduce data occupation. Then, a k-Nearest Neighbours algorithm classifies the samples, achieving as low as 1.56\% of misclassification. 
Despite the optimal results achieved, these work rely on a set of hand-tuned parameters, which impair the generality of the model over time. To retrain these parameters, these models need the entire history of the data, which (i) is not always available, and (ii) causes the system to necessitate of a single cloud orchestrating unit.

The most recent study based on data modeling is~\cite{ref2-46}, which takes advantage of real-case vibration data of a bridge in China and datasets from laboratory structures. They propose an approach called "in-network damage detection on edge" to detect bridge structure damage. 
They collect statistical features of the input data into an \textit{m}-dimensional feature vector.
Then, they fit a Gaussian distribution model on the training set and consider anomalies as the tail of this distribution.
Although the trained model's accuracy on the recorded Yonghe Bridge in China reaches 100\%, it decreases to 96\% for the secondary simulated structure case. Furthermore, for all the experiments in this work, accuracy varies between 85\%-100\%. This high fluctuation in performance is due to the reduced number of extracted features which impair the capability of this approach to model the structure's behavior in different positions with several sensors. While the lack of adjustability to the structure's behavior over time is a limit of their work, we propose a solution to update the model adaptively after a change in behavior has been observed.    
Santos et al.~\cite{ref2-42} is the only approach fully deployed on the edge. It computes the Fast Fourier Transformation (FFT) of input vibration data and the difference in peak frequency of each consecutive 0.5 second time window at the node. Then, computed natural frequencies are sent to sensors heads (i.e., Gateway) to estimate the structure's status using a threshold-based algorithm which results in a perfect damage detection with only 2 False Positives (FP). Transmitting only natural frequencies causes a network traffic of 6.720 KB/h. In a similar vein, we further decrease network traffic to only 10B/h by applying the Principal Component Analysis for data compression and classification directly on the node.
Noteworthy, as demonstrated in section \ref{sec:software_results}, employing frequency features strongly impairs anomaly detection performance on our structure, making this approach unsuitable for our problem.
Similar to Santos et al.~\cite{ref2-42}, other works, e.g.,~\cite{added-ref2-1,added-ref2-2,added-ref-2-3}, study the pros and cons of cloud computing and edge computing in the content of SHM systems.         

To the best of our knowledge, all of the statistical data-modeling approaches exploited expensive piezoelectric accelerometer to collect data. In contrast, we replace such sensors with a low-cost, low-power (but higher noise) MEMS accelerometers.
\subsubsection{Deep Neural Networks}
In~\cite{ref2-43, ref2-44}, the authors present two DNN-based approaches. 
A 1D-CNN is used in \cite{ref2-43} to estimate the Probability of Damage (PoD) on the BM benchmark \cite{dyke2003experimental}. A PoD close to 0 points to the normal case, whereas a PoD of 1 corresponds to the damaged condition. Evaluating nine scenarios of increasing damage severity shows that their 1D-CNN correctly ranks the scenarios from one to nine by correctly predicting an increasing damage condition.
Compared to the conventional 3D CNNs, 1D CNNs require less computational complexity, thus take less time to train the model. 
However, this model still requires data generated by a cluster of nodes, not a single node, to achieve high accuracy, which is unsuitable for an online-training on edge-nodes.
On a totally different input data, images, Wu et al.~\cite{ref2-45} present an approach for online inference. The authors exploit two deep convolutional neural networks, namely VGG16 and ResNet18, for crack and corrosion detection of structures from image data. 
They apply aggressive pruning to reduce the complexity while maintaining a high detection accuracy (they reduce VGG16 memory footprint to 44 MB and ResNet18 to 2MB). 
Running the algorithms on a Jetson TX2 platform, the authors achieve 94.6\%-98.5\% detection accuracy for crack detection on different nuclear power plant structures and 82.8\% detection accuracy in corrosion images of different metallic surfaces. 
Despite the performance, we do not employ deep supervised neural networks since they require a large training labeled dataset that is unavailable in our case, and, more in general, labeled anomaly data is not available in typical structural health monitoring installations.
\subsubsection{Data Reduction}
The last category of works exploits compression algorithms for damage detection. These algorithms first compress and reconstruct the input data and then compute the difference between original and reconstructed signals. The higher is the difference between the original and the reconstructed signals, the higher is the probability of damage.
In this context, autoencoder (AE) neural networks are among the most popular approaches. 
For example,~\cite{ref2-29} uses an AE for damage diagnosis on a laboratory's synthetic bridge for indirect bridge monitoring scenarios, outperforming all other anomaly detection algorithms with MSE $\approx 5$ in computing 30 levels of damage severity.
Given the promising performance of the method, we also test autoencoder-based anomaly detection in our work.
Furthermore, linear processing-based compression methods such as principal component analysis (PCA) also achieve good performance in SHM for damage detection (e.g.,~\cite{ref2-30, ref2-31, ref2-32, ref2-33}). 
For example,~\cite{ref2-31} describes moving PCA on vibration data. They show the effectiveness of compression by evaluating the model over a laboratory beam bridge and recorded data of a bridge of Guangdong, China, with 100\% damage identification. 
Even though the works mentioned above can reach perfect accuracy, training and inferring are performances on unconstrained remote devices (e.g., i7 intel processor) after data transmission and collection.
In our work, we aim to tackle these models' generalizability and deployability by introducing a new lightweight pipeline. Compared to other SHM works, we propose a method to constantly update the anomaly detector, tackling the time variability of the structure dynamic over time; also, our approach entirely relies on unsupervised data, not necessitating for labels as other DNN-based approaches.
Finally, to the best of our knowledge, we are the first to deploy and analyze the performance of a complete anomaly detection pipeline on an in-situ sensor network utilizing real-life SHM system installation on a viaduct.
\begin{comment}

\begin{itemize}
    \item Our in-field collected dataset from a real-life SHM installation on a viaduct in Italy is an excellent proxy for moderate modification in the structure's behavior, indicating natural perturbations like wind and earthquake. 
    \item Our unsupervised training methodology, whose indispensability is undeniable for SHM systems i.e., labelled damage data is absent in majority of the cases, cannot be easily achieved by DNNs or statistical data modeling in former vibration-based SHM systems. 
    \item Avoiding hand-crafted feature extraction of statistical data models and complexity of DNNs grant model's adaption at the occurrence of slight modifications due to wind or earthquake.      
    \item Node implementation of the solution provides a quantized analysis of memory footprint, computational load, and power consumption. Furthermore, bringing the decision unit of the SHM system to the node reduces network traffic to 10B/h. Thus, this reduction cancels comprehensive data collection from the sensor and enables a scalable solution for the large structures.   
\end{itemize}
\end{comment}

%/
\section{SHM Installation \& Background}
\label{sec:bakground}
\subsection{Bridge Structure}
\label{subsec:bridge}

The vibration data analyzed in this work comes from a viaduct located in northern Italy on the ss335 state highway, composed of 18 different sections. 
Last year,  the viaduct underwent a technical intervention to strengthen the structure of a viaduct section, with a corresponding change in the vibration signal produced by its structure.
Before the intervention, this section has been instrumented with five SHM nodes to monitor viaduct vibration, as illustrated in Fig \ref{fig:SHMframework}. 
For this reason, in this work, we analyze the unique situation of accelerations gathered before and after this strengthen intervention.
We use these data as a proxy for an abrupt change in the viaduct structure caused, for example, by external factors such as an earthquake.
After the intervention, we consider the vibrations raw data as the normal data produced by a 'sane' viaduct. Conversely, the accelerations measured before the intervention are used as the 'anomaly', given the high degradation of the bridge's structure.
\subsection{SHM Network}
\label{subsec:shmframework}
\begin{figure}
  \centering
\includegraphics[width=0.9\columnwidth]{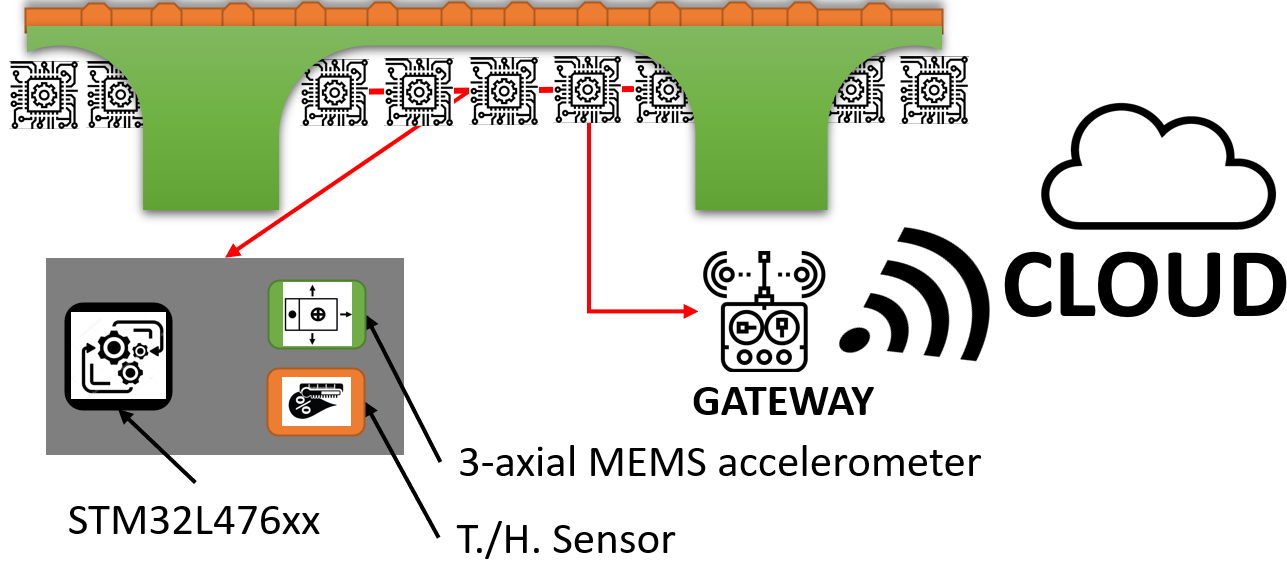}
  \caption{Overview of the installed monitoring system on the viaduct. Five sensors are linked to a gateway for streaming data to the cloud. The grey box showcases the main components of a sensor node.}
\label{fig:SHMframework}
 \vspace{-0.5cm}
\end{figure}
The depicted installation is a vibration-based SHM system, which exploits acceleration gathered from the sensors to detect damages and monitor the viaduct health condition.
Fig. \ref{fig:SHMframework} shows the installation composed of five nodes connected via CAN-BUS to transmit data to a gateway: in the baseline setting, no computation is performed neither on the nodes nor the gateway.
Nodes gather and transmit data to the gateway. The gateway sends the sensor's data to the cloud for storage purposes. All the processing on the data is then carried out daily on the cloud.

The gateway is a Raspberry Pi 3 module B (RPi3), an edge computer actively employed in many fields such as robotics, smart sensors, or SHM. It includes a Broadcom BCM2837 SoC, with 64 bits 4-core Cortex-A53 running at 1.2 GHz and 1 GB of DDR2 RAM. The gateway supports Linux operating system, allowing for typical python applications deployment for either communication (e.g., an MQTT broker \cite{mqtt}) and in-field machine learning (e.g., Keras \cite{keras}, scikit-learn \cite{sikit}).

The sensor node is represented in the lower part of Fig. \ref{fig:SHMframework}.
It is composed of the LIS344ALH analog tri-axial accelerometer \cite{Lis344}, the HTS221 temperature and humidity sensor \cite{hts221}, and an STM32L476VGTx microcontroller as a computational core. 
The core features a floating-point unit and a digital signal processing (DSP) library, which has been used in our work to optimize the algorithm deployment.
The microcontroller unit is an ARM 32-bit Cortex-M4 running at 80 MHz, with 96 KB of SRAM and 1MB of Flash memory.
This node samples the acceleration with the internal ADC at a frequency of $25.6$ kHz. 
For increasing the bit-resolution, windows of 256 samples are filtered with a 6-state FIR filter and reduced to a single value, thus obtaining a final sampling rate of 100 Hz. 

\begin{figure*}
  \centering
\includegraphics[width=1\textwidth]{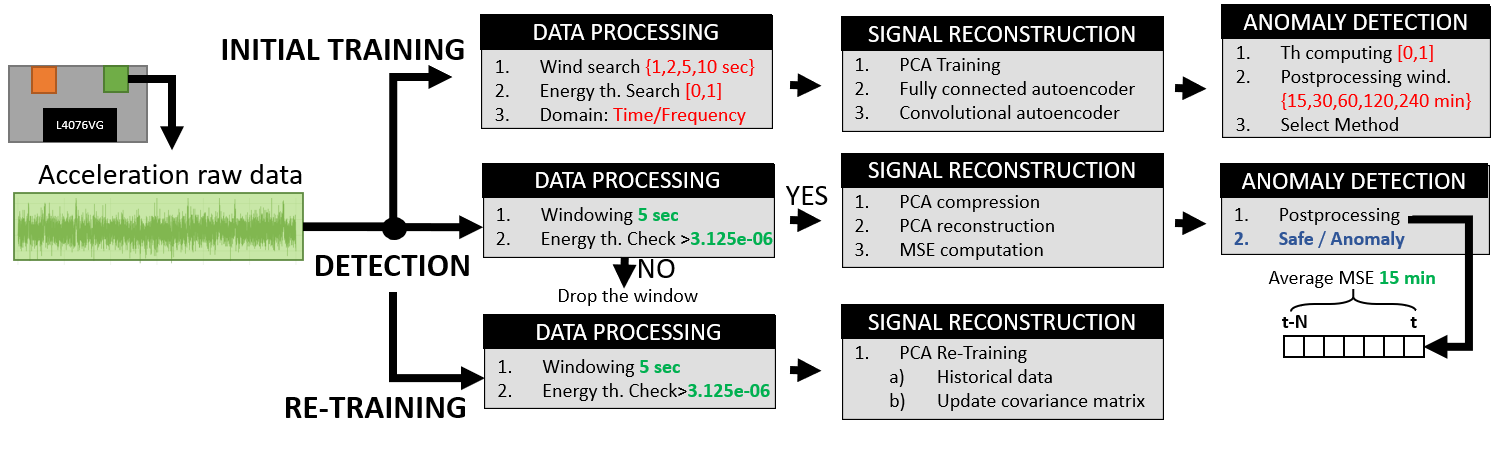}
  \caption{The proposed framework to analyze the condition of a viaduct starting from raw acceleration data. In the top part of the figure, we show the hyper-parameter tuning (in red) and the initial training steps done before the first time that the monitoring system is activated.
  In the middle, we show the inference steps to be done continuously for safe/anomaly condition assessment. In the bottom part, we show the possibility of updating the signal reconstruction algorithms after the pipeline detects an anomalous event to avoid the increase of false-positive alarms due to the bridge’s static deformation caused by wind or ageing.}
\label{fig:pipeline}
 \vspace{-0.3cm}
\end{figure*}
\subsection{{Anomaly detection approaches}}
\label{subsec:Math}
\subsubsection{{Principal Component Analysis}}
Principal Component Analysis (PCA) is a method to deal with high dimensional correlated data by transforming them into minimally correlated data \cite{ref2-49}. Exploiting the covariance matrix of high dimensional data, the PCA projects it into a new space where the axes correspond to the eigenvectors of the covariance matrix, ordered by the value of their eigenvalues. 
PCA reduces data size by preserving only directions that retain most of the information \cite{ref2-50} (the ones with the associated higher eigenvalues). 
Considering a $M\times N$ dimensional data matrix 
$x = \begin{bmatrix}
 x_1, & x_2,& x_3, & ... &, x_N
\end{bmatrix}$
where $x_k$ is a column vector of $M$ features representing a sample, its normalized covariance matrix is
\begin{equation}
\Sigma = \frac{1}{N-1} \displaystyle\sum_{k=0}^N {(x_k - \Tilde{x})(x_k - \Tilde{x})^T } 
\end{equation}
where $\Sigma$ is a square $M\times M$ matrix. Its diagonal holds variance of each individual sample, and off-diagonal values are covariances of sample combinations. Using eigenvalue decomposition, we can write
 \begin{equation}
 \Sigma = V \Lambda V^{-1}.    
 \end{equation}
where $V$ columns represents the eigenvectors, and the principal diagonal of $\Lambda$ contains corresponding eigenvalues.
It can be proven that $\mathbf{V}^k\in\mathbb{R}^{k}$ is a basis of the sub-space of dimensions $\mathbb{R}^{k}$ which retains the highest similarity with the original one.

Due to the usually high number of features, $M$, PCA requires a high memory footprint to store and compute the covariance $M\times M$ matrix and extract its eigenvalues.
To cope with this limitation on low-memory edge devices, the authors of \cite{ref3-4} introduced \textit{History PCA}, a streaming algorithm to train the PCA without storing data, which has been later deployed on edge/nodes by \cite{Ref3-5}.
Compared to other streaming approaches, HPCA exploits the history of the data and new samples to update the partial covariance matrix, allowing a faster convergence and better accuracy \cite{Ref3-5, ref3-4}.
In our work, we exploit HPCA to deploy our algorithm on the sensor nodes, moving both training and detection from the gateway to the leaf nodes of the SHM sensor network.

\subsubsection{Autoencoders} 
\label{subsubsec:autoencoders}
Autoencoders are neural networks composed of two or more layers used to compress data and detect anomalies \cite{ref3-14}.
Autoencoders can be segmented into two parts, Encoder and Decoder. 
The encoder, $f_E(x)$, projects the input data $x \in \mathbb{R}^{M}$ into a lower-dimensions hidden space $h \in \mathbb{R}^{k}$, exploiting one or multiple layers, either fully connected, convolutional or recurrent \cite{ref3-13}.
An example of a single-layer encoder is
\begin{equation}
    h = f_E(x) = \Phi (W_E x + b_E)  
\end{equation}  
where $W$ is the weight matrix, and $\Phi$ is the activation function of a single layer. 
The decoder $f_D(h)$ projects back the compressed signal $h$ to its original space, creating a new signal $\bar{x}\in \mathbb{R}^{M}$ as 
\begin{equation}
    \bar{x} = f_D(h) = \Phi (W_D h + b_D).  
\end{equation}  

The model's training favours the similarity of $x$ and $\bar{x}$ without employing data labels, teaching the encoder to find the best-hidden space that mainly preserves the features of the original one.
During training, the loss function is represented by a similarity metric between the original and the reconstructed signal.

The same metrics are also exploited to employ the autoencoder as an anomaly detector.
Reconstructed signals, similar to those encountered during training, result in a low reconstruction error. On the other hand, reconstructing signals with different characteristics than those used for training are badly reconstructed.
To detect anomalies, only normal signals are fed to the autoencoder for training.
Therefore, new anomalous signals encountered during the test phase are poorly reconstructed, with a higher mean square error (MSE), and thus identified as anomalies.

\section{Methods: Anomaly Detection in an SHM framework} 
\label{sec:scenarios}
This section describes the main contribution of this work. We discuss our novel SHM pipeline and its deployment to augment the SHM installations to raise \textsl{integrity} alarms automatically.
We first show our complete pipeline, comprising a step of initial training, the in-field estimation, and the possibility for an online update of the models.
Then, we describe our proposed solutions to efficiently embed our pipeline inside the existing system, reducing the energy consumption and network traffic while maximizing the system's scalability for large SHM installations.
\subsection{Anomaly detection pipeline}
\label{subsec:anomaly_detection_pipeline}
As shown in Fig. \ref{fig:pipeline}, our pipeline is composed of three main blocks (from left to right in the figure).
First, a series of transformations such as windowing, data filtering, and feature extraction is applied.
Then, the signal compression-decompression algorithm for anomaly detection is applied. We tested three algorithms: i) PCA, ii) a Fully-connected autoencoder, and iii) a convolutional autoencoder.
Finally, the MSE between decompressed and original signal is computed to detect the structural integrity of the viaduct. An average over time is calculated to smooth the damage detection, reducing false alarms.

\subsubsection{Pre-processing}  
\label{subsubsec:pre-processing}
\begin{figure*}
  \centering
\includegraphics[width=1.0\textwidth]{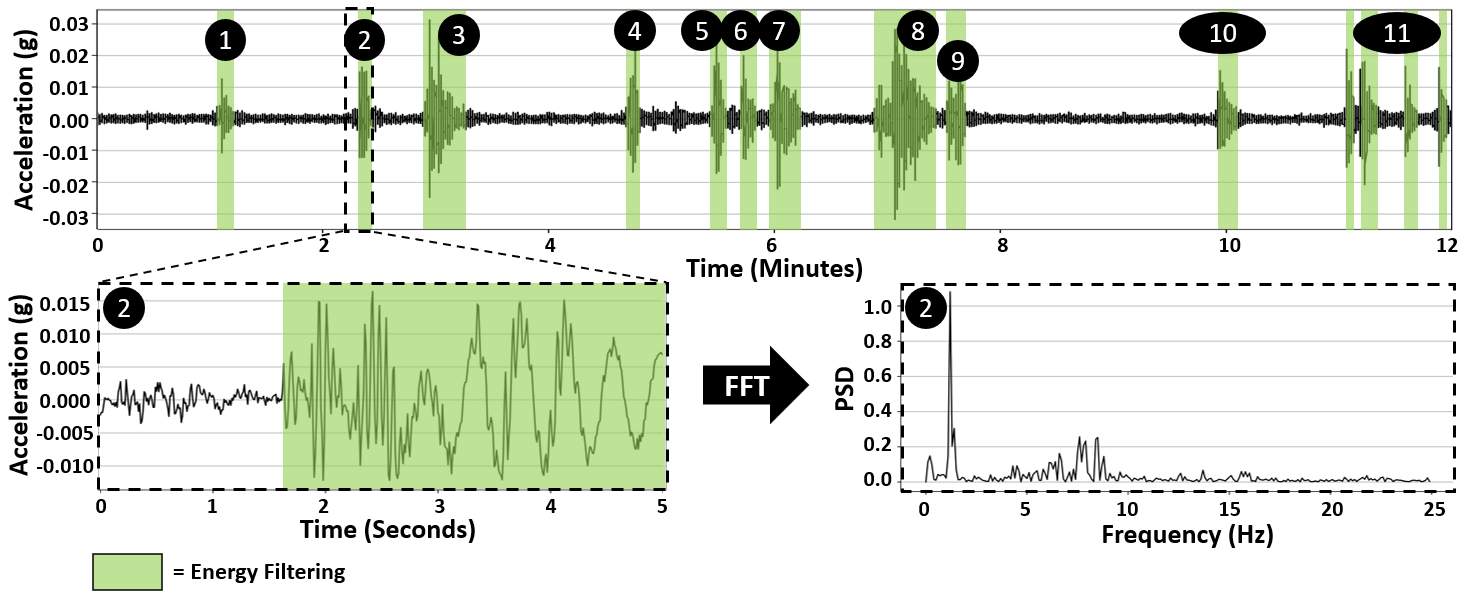}
  \caption{Top panel. Twelve minutes of mean-centered raw acceleration data of the $z-$ axis of the middle sensor installed on a pier of the bridge. Peaks are associated with vehicle passages. Left panel. Zoom of a 5-second window containing the oscillation associated with the passage of a vehicle. Right panel. Frequency response of the window of signal highlighted with the dashed rectangle.}
\label{fig:acc_data}
 \vspace{-0.5cm}
\end{figure*}

This step covers the windowing of raw signal, the energy extraction, and eventually the application of the FFT, if needed.
We used a single acceleration axis for our analysis, namely the $z$-axis (i.e., vertical axis) of the sensor installed in the middle of the section since it contains the most critical information about the bridge.

As Fig. \ref{fig:pipeline} shows, data processing starts by dividing acceleration raw data into non-overlapping windows, similar to~\cite{Ref3-5}.
We explore window dimensions of 1 to 10 seconds to balance accuracy with algorithm complexity.
Noteworthy, given the hardware-related constraints such as limited memory and hard time constraints, different window dimensions can fit different use-cases.

After, we check the energy of the windowed signal. 
In our case, the analyzed bridge does not experience heavy traffic, hence it is often resulting in low vibration windows, containing only the white noise of the sensor.
Therefore, we designed an energy-based window cleaning to eliminate non-informative windows. 
To this end, the energy of each window is extracted and compared to a trained energy threshold.
Windows with an energy lower than the trained threshold are removed from further analysis.  
Energy of each window is computed as follows: 
\begin{equation}
E = \displaystyle\sum_{i=1}^{W_d} X_i^2 
\end{equation}
where $W_d$ is the width of each window.
The search of energy threshold is done exploiting the iterative steps of Alg. \ref{alg:energy}. 
At each step, an increase in the threshold leads to removing a higher percentage of the windows.    
Alg. \ref{alg:energy} stops when the reconstructed signal of not filtered-out windows drops below a predetermined Quality of Service (QoS), namely the average reconstructed signal-to-noise ratio (RSNR), computed as $\RSNR = 20 \log_{10} \left( \frac{\norm{\mathbf{x}}_2}{\norm{\mathbf{x} - \hat{\mathbf{x}}}_2} \right)$, with $\mathbf{x}$, the original signal, and $\hat{\mathbf{x}}$, the reconstructed one.  
Based on \cite{Ref3-5} and considering a compression factor of $15\times$ as in \cite{Ref3-5}, we set this lower bound average RSNR to 16 dB.
 \begin{algorithm}
 \begin{algorithmic}[1]
 \caption{\label{alg:energy} Energy Filtering}
 \State Input: ${\mathbf{X_{train}, X_{val}}}$
 \State $th = 10^{-10}$
 \Do
    \State $th += 2^{-8}$
    \State $\mathbf{X_{train}, X_{val}}$ $\gets$ filter ($\mathbf{X_{train}, X_{val}}, th$)
     \State $W \gets pca(\mathbf{X_{train}})$
     \State $X_r \gets \mathbf{X_{val}WW^\intercal}$
     \State $S \gets RSNR(\mathbf{X, X_r})$
 \doWhile{$S$ $<$ 16 dB}
 \State Output: $th$
 \end{algorithmic}
 \end{algorithm}
%                   End of Blanks Space
Fig. \ref{fig:acc_data} shows acceleration data and highlights the portion of the signals selected by the tuned energy threshold with green background.
Fig.\ref{fig:acc_data}-B and Fig.\ref{fig:acc_data}-C  show a zoom of peak \circled{2} in the time and frequency domain.
As detailed in Sec. \ref{sec:software_results}, applying this energy filtering improves the accuracy of all our analyses.  
\subsubsection{Signal Reconstruction}
\label{subsubsec:signal_reconstruction}
We process the non-discarded windows with different compression-decompression models. 
The similarity of the original with the reconstructed signal is then used to detect anomalies.  

%                   End of Blanks Space
This phase is split into two steps: i) compression and ii) reconstruction of input pre-processed signals.
We test one model-driven method, namely the PCA, and two data-driven approaches, a fully connected autoencoder and a convolutional autoencoder, as anomaly detectors.  
We impose a compression factor of the input signal of $16 \times$, before reconstruction. 
In PCA, we keep the top 16 principal components. In the fully connected autoencoder, we employed 16 neurons in the hidden layer. In the convolutional autoencoder, we utilized a stride over convolutional layers of the encoder part of 32, reducing from 500 to 16 the dimension of the signal before the transposed convolutions. 
The PCA and fully connected autoencoder perform the same number and type of operations (two matrix multiplications, $\mathbf{A}$ $\times$ $\mathbf{B}$, and $\mathbf{B}$ $\times$ $\mathbf{C}$, with dimensions $\mathbf{A}$ 1$\times$500, $\mathbf{B}$ 500$\times$16 and $\mathbf{C}$ 16$\times$500 ) and only differ in the training approaches: the first one is model-based, while the second is trained via a data-driven back-propagation.  
The convolution autoencoder is composed of 8 hidden layers followed by ReLU activations. Adam optimizer, along with 80 epochs, is used to train this model.  
\subsubsection{Anomaly Detection}
\label{subsubsec:anomalyDetection}
\begin{figure}
  \centering
\includegraphics[width=1\columnwidth]{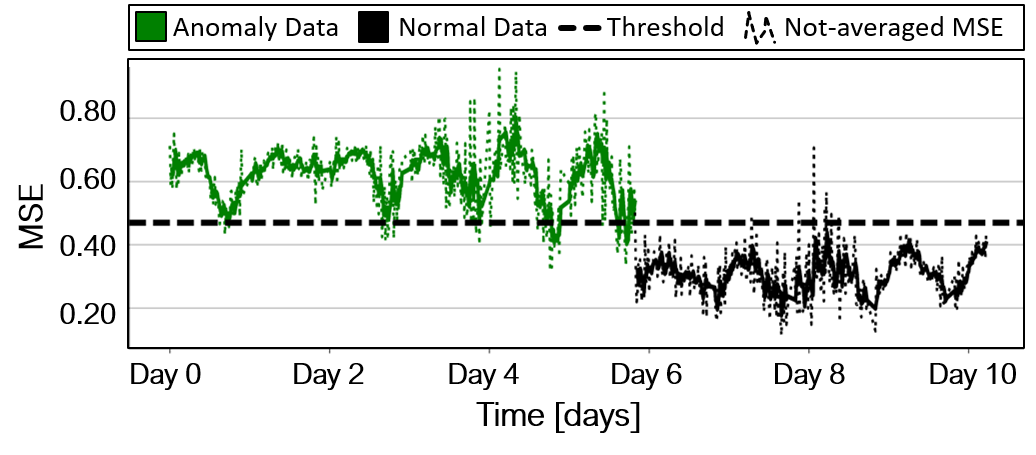}
  \caption{PCA output mean square error (MSE) on the test dataset. Input window dimension is set to 5 seconds. Solid line is obtained by applying the post-processing with window dimension = 1 hour.}
\label{fig:mean_square_res}
\vspace{-0.5cm}
\end{figure}
We use the difference between the original and reconstructed signals as an anomaly detection score. A higher difference implies a worse reconstruction.
In particular, we compute the mean square error (MSE) as:
\begin{equation}
MSE = ||x_i - \Bar{x_i}||_{L2}=\displaystyle\frac{1}{n}\displaystyle\sum_{i=0}^{n-1} (x_i - \Bar{x_i})^2 
\end{equation}
where $x$ is the original signal, $\Bar{x}$ is the reconstructed signal, and $n$ is the number of samples in a window.

Our reconstruction algorithms are trained solely with normal data using an unsupervised process.
Therefore, the algorithms should reconstruct normal data with low MSE, while they cannot reconstruct anomalies that show a different signal dynamic not seen during the training, leading to a higher MSE. 
A threshold to distinguish the two data classes can be thus statistically derived solely by normal validation data. 
To compute it, in our case, we compress a validation set of normal data using the different compression algorithms. Then, we select the threshold as the mean of the MSE over all the data compressed plus three times its standard deviation ($th = \mu + 3\times\sigma$). 
Noteworthy, we set this threshold to have only $0.01\%$ of statistical false positive errors. 
The results of this procedure are shown in Fig \ref{fig:mean_square_res}, where PCA is used to compute MSE over normal and abnormal data. 

To further reduce the false alarms, we propose an average over time of the soft-predictions (MSE values).
We explore windows between 15 minutes to 4 hours, showing that a larger window is positively correlated with better accuracy, increasing the gap between reconstructed normal data and reconstructed anomalies but causes larger delays in prediction.
\subsection{Algorithm Phases: Train, Detect, Re-Train} 
\label{subsec:algPhases}
Our pipeline is characterized by three main phases (Fig. \ref{fig:pipeline}, top-down), namely i) an initial algorithm selection, parameter tuning, and model training, ii) the continuous bridge monitoring, and iii) a re-training phase to adapt the model to slow modifications of the bridge dynamic.
The first phase, \textit{training}, begins with an ablation study over the possible hyper-parameters: the input window dimension, the tuning of the energy filtering step, the anomaly detection models parameters, and the post-processing. 
After the definition of the parameters, the chosen model is trained with the normal data of the viaduct.

The second phase, \textit{continuous monitoring}, exploits the best solution found during the training to perform a long-term online detection of the viaduct damages.

The last phase, \textit{re-training}, involves the update of the model parameters over time to adapt to the temporal-changing dynamic of the signal. This step is primary for this kind of analysis since modal analysis shows that light stresses such as wind or traffic load cause slow structural modifications, resulting in slightly different signal dynamics.
Further, in SHM scenarios, false alarms can not be tolerated since they can trigger critical alarms causing a bridge maintenance intervention with a consequently high cost.

\subsection{Deployment: Sensor Vs. Cloud}
Deploying our proposed anomaly detection pipeline (Fig. \ref{fig:pipeline}) is not trivial due to problems such as the scalability in the number of nodes or the lifetime of the nodes.
Data communication costs become critical when multiple streams must be transmitted to the cloud. At the same time, the limited memory footprint of tiny edge devices is a major constraint for on-sensor computing.
Therefore, we here discuss three deployment scenarios of our anomaly detection pipeline on our SHM system composed of the sensor network installed on the viaduct and the cloud that augments the system with data storage and computation capabilities.
Specifically, we discuss the trade-offs of performing the three different phases (i.e., training, anomaly detection, and re-training) of our algorithm either on the cloud or on the nodes, or as a mix of them.
Noteworthy, these reasonings also hold for much bigger SHM installation of hundreds of nodes, where the scalability issues and the data storage can be the real bottleneck of the system.
\subsubsection{Cloud Computing}
\label{subsec:cloud}
\begin{figure}
  \centering
\includegraphics[width=1\columnwidth]{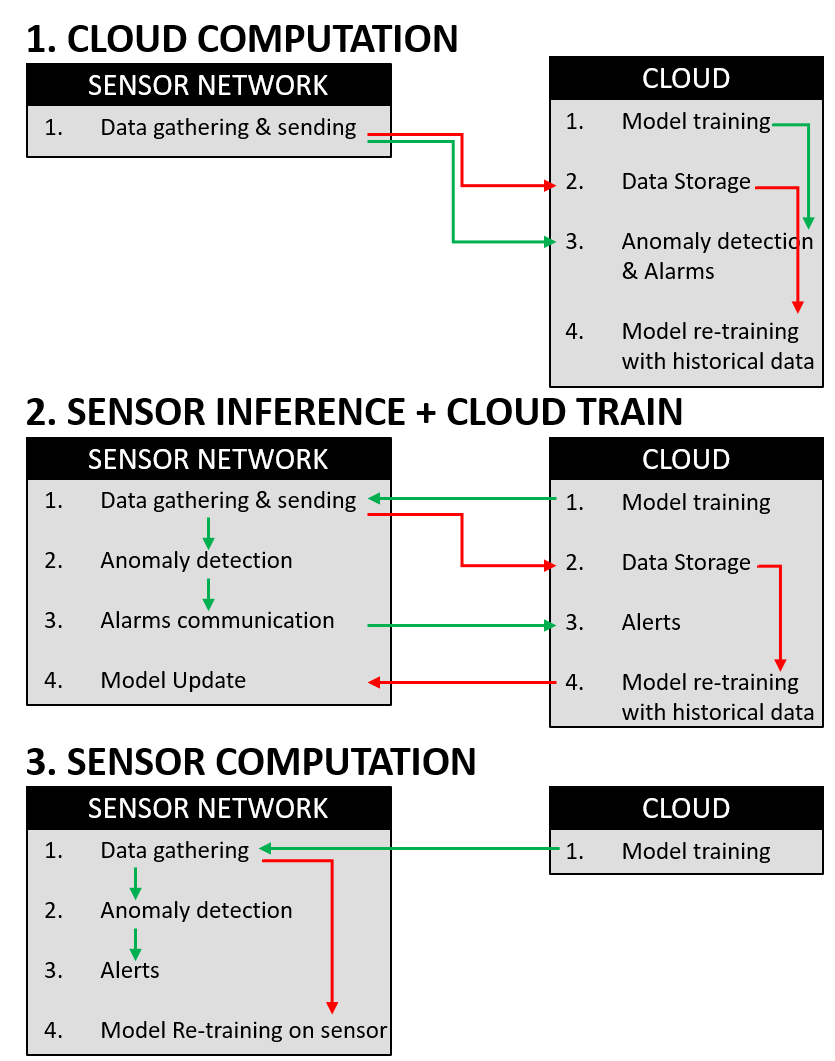}
  \caption{Three deployment scenarios of our anomaly detection pipeline. Green arrows highlights the inference steps, while red ones the re-training and updating of the model over time.}
\label{fig:scenarios}
 \vspace{-0.5cm}
\end{figure}
As shown in Fig. \ref{fig:scenarios}-1, transmitting all the data to the cloud while having no processing in the sensor network causes i) a high data communication cost, ii) the necessity of cloud data storage, and iii) a daily cloud computation of anomalies, alarms, and, less frequently, the iv) re-training of the model.
The transmission of data to the cloud is the first issue in this scenario.
Although several cost minimization techniques such as new communication paradigms \cite{Ref3-10, Ref3-11} or edge data-reduction \cite{Ref3-5} have been introduced recently, data communication still represents the highest installation cost over months in terms of energy. 
Using one of the most efficient standard protocol stacks available today, the Narrow Band Internet of Things (NB-IoT) \cite{Ref3-9}, which has demonstrated optimal performance in the SHM field, the system consumes up to 0.94 J for a typical transmission of 500 bytes in the open space, decreasing the maximum lifetime of the SHM nodes and thus needing solutions such as energy harvesting \cite{ref3-15} or a wired sensor.
Furthermore, the different cloud service providers such as Amazon, Microsoft, and Google account for data computation costs as pay-to-go, with the client paying for the computational time exploited \cite{Ref3-12}, also making the money invested in this service not negligible. 
Therefore, a complete cloud paradigm for anomaly detection causes a higher maintenance cost and shortens the lifetime of the SHM nodes, demanding more frequent interventions on the installation.
\subsubsection{Sensor Interface with Cloud}
Involving sensors in the computation reduces the anomaly detection pipeline's costs. 
The anomaly detection model is exported to the sensor to predict the viaduct behaviour, while the model re-training is still performed in the cloud.
Fig \ref{fig:scenarios}-2 shows the overall functionality of this approach. In green, we highlight the anomaly detection pipeline, while in red, the update of the model over time. 
Note that while we can reduce both the traffic (streaming only data when we decide to start a re-training) and the cloud computation (only the re-training function is executed on the cloud),  cloud storage and processing cost still remain an issue for this kind of scenario, making the scalability an open problem in this kind of approaches.

In our use case, we deploy the anomaly detection pipeline on the node for this scenario; while keeping the data streaming to the cloud for algorithm re-training.
After the on-cloud algorithm re-training, the new model is deployed on the nodes.

\subsubsection{Sensor Computing}
To also eliminate the communication costs for re-training, we propose to move both the computation of the online anomaly detection and the update of the node's model on the sensor.
Using this approach, after the initial training, done one time per SHM installation, no further computation is required from the cloud. Each SHM installation can be considered a standalone unit without the need for cloud communication unless an anomaly is detected.
In this scenario, the scalability is not more a problem since the cloud is only used to monitor the sensors' status and initialize them.
Fig. \ref{fig:scenarios}-3 highlights the steps of this approach. 

For our use case, while the porting of the anomaly detector is trivial, training PCA on a memory-constrained device entails many challenges, such as storing the covariance matrix in a memory constraint microcontroller.
Further, storing many data on local nodes is impossible, given the low FLASH memory. 
Thus, we employ streaming PCA, previously deployed on a sensor node in \cite{Ref3-5}, which aims at finding a compression matrix sequentially to avoid i) storing lots of data at edge and ii) compute the entire covariance matrix \cite{ref3-4}. 
Compared to \cite{Ref3-5}, instead of employing the PCA only for data compression, we use it also to perform anomaly detection at the edge of the sensor network.

\section{Experimental Results: algorithm exploration} 
\label{sec:software_results}
In this section, we mainly focus on analyzing the proposed framework in Fig. \ref{fig:pipeline}. Using grid search over different hyperparameters and framework elements, we explore the performance of our pipeline while changing both its blocks (e.g., anomaly detection techniques) and the block's parameters (e.g., by presence or absence of the energy filtering while tuning its threshold).
After, we examine our best detector's robustness, artificially changing the severity of the anomaly in the dataset and correlating severity with algorithm performance.
Finally, we compare the proposed anomaly detectors with state-of-the-art ones. To be fair, we reproduced the state-of-the-art algorithms and we applied them to our data using the same input window dimension and post-processing. 
\subsection{Notations \& Benchmark}
\label{subsec:benchmark}
First, we introduce the notations and metrics that we use to evaluate the different methods and hyperparameters in this work.
We use three metrics for performance assessment:\\
i) accuracy, the total correctly classified windows
$$
Acc. = \frac{TP + TN}{TP + FP + TN + FN}
$$
ii) sensitivity, the percentage of correctly detected anomalies
$$
Sens. = \frac{TP}{P} = \frac{TP}{TP + FN}
$$
iii) specificity, the percentage of correctly classified normal windows 
$$
Spec. = \frac{TN}{N} = \frac{TN}{FP + TN}
$$
Where $P$ are the positives, $N$ the negatives, $TP$ are the true positives, $TN$ are the true negatives, $FP$ are the false positives, and $FN$ are the false negatives.
Furthermore, we use Area Under Curve (AUC) to assess the performance of our models. 
For our purpose, we consider the "anomalies" as positives prior to the intervention, while the negatives are windows of "normal" data after the intervention.  
With Compression Factor (CF), we point to the ratio between high-dimensional original space and algorithms reduced data space, i.e., the projected PCA data and the latent autoencoders data. 
Finally, we define input dimension, the length of each non-overlapping window in the data processing step, and output dimension, the total time considered after averaging multiple windows before final classification.

Our test dataset comprises 25 days of continuous monitoring of the viaduct with 5 sensors described in Sec. \ref{subsec:bridge}.
For our analysis, we consider the central sensor of the chain, which is most influenced by the viaduct vibration. 
Note that using a higher number of sensors does not improve the accuracy in this case, but it is still feasible.
The data are composed of 5 days before the maintenance intervention, labeled as anomalies, and 20 days after, labeled as normal data.
We select as the test set all the 5 days of anomalies and 5 days of normal data to have a balanced test dataset.
We divided the remaining 15 days of normal data into a validation set (5 days) and the training set (10 days).    
Note that anomalies are used neither in training nor validation datasets, given that all analyses are unsupervised. The anomalies are used in our results only to assess the classification accuracy of our approaches.

To the best of our knowledge, considering the viaduct's unique condition, this is the first anomaly labeled data from a real-life viaduct.
\subsection{Model selection \& Data domain}
\label{subsec:method_comparison}

\begin{table}
\centering
\caption{Performance of our pipeline changing anomaly detection algorithm with discrete wavelet transform, frequency and time data as input space domain.}
\label{tab:accuracies}
\begin{tabular}{|l|l|lll|}
\hline
Algorithm       & Domain & Acc. & Spec. & Sens.  \\\hline
\multirow{2}{*}{PCA}                 & Raw        & 98.8 \% & 100   \%& 97.33\%  \\
                                     & FFT        & 77.22 \% & 99.20  \%& 50.63 \%\\
                                     & DWT        & 84.36 \%  & 96.79 \%    & 74.44 \%  \\\hline
\multirow{2}{*}{FC Autoencoder}     & Raw        & 68.75 \% & 99.73  \%& 44.04 \% \\
                                     & FFT        & 56.02 \% & 96.54  \%& 23.61 \%\\ 
                                     & DWT        & 69.99 \% & 97.87 \%   & 47.66 \% \\\hline
\multirow{2}{*}{Conv. Autoencoder}    & Raw        & 50.6 \% & 67.2  \%& 37.1 \% \\
                                     & FFT        & 56.30 \% & 85.28  \%& 32.66\%\\
                                     & DWT        & 52.12 \% & 100 \%   & 13.83 \% \\\hline
\end{tabular}
\end{table}

\begin{figure}
  \centering
\includegraphics[scale=0.55]{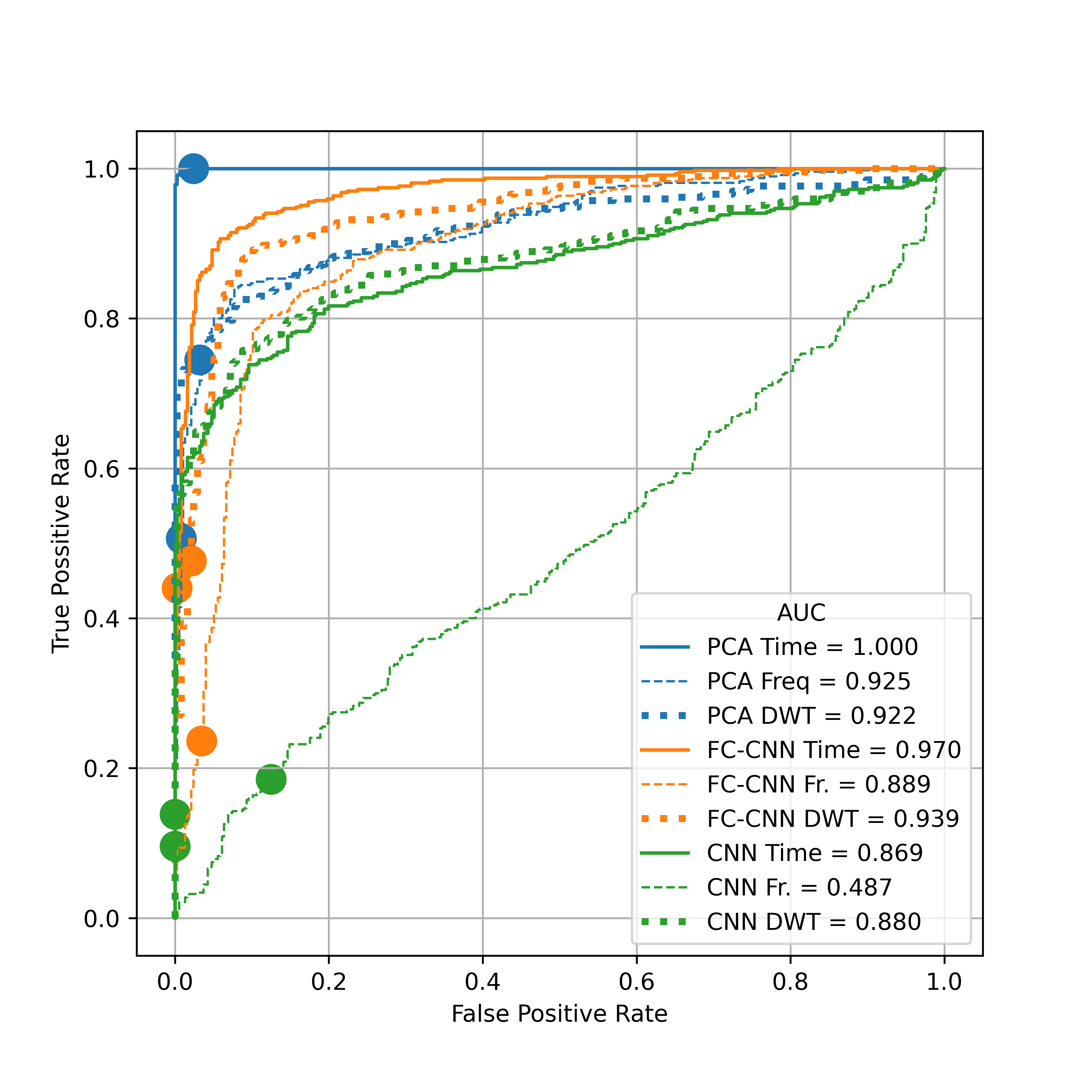}
  \caption{ROC curve for different input signal domains, i.e., time, frequency, and time-frequency.}
\label{fig:roc}
 \vspace{-0.5cm}
\end{figure}
Modal analysis is the gold standard used to analyze the dynamic characteristics of large-scale buildings~\cite{ref3-16}. 
On the other hand, previous studies have demonstrated the feasibility of using raw time series for anomaly detection \cite{Ref3-5}.
Hence, both time and frequency domains are promising directions to analyze. 
Therefore, we test three anomaly detectors fed with both frequency and time inputs.
We selected PCA and Autoencoders as detectors given their already demonstrated success in anomaly detection and, more precisely, on SHM tasks \cite{ref3-18}.
We fix the input window dimension to 5 seconds of accelerometer output samples and the output dimension to 60 minutes for this comparison.
The compression factor is fixed to 16; therefore, we select the most significant 16 principal components for PCA, while ensuring the innermost latent dimension of both fully connected and convolutional autoencoders to have a dimension of 16.   

Table \ref{tab:accuracies} and Fig. \ref{fig:roc} report the evaluation results on the three models using the 10 days of the test set, using both the input data domains.
As previously described, to report accuracy, sensibility, and specificity, we use a threshold on the output MSE of $\mu + 3\times\sigma$. On the other hand, the Receiver Operating Characteristics (ROC) curve are threshold independents.
Using time-domain input, PCA outperforms both the other two models reaching  $98.8 \%$, $100 \%$ and $97.33 \%$ of accuracy, specificity, and sensitivity, respectively, and an approximate 1.00 AUC. 
Notice that PCA is the only method to remove all the false alarms in the system, preventing sending false alarms to bridge maintainers. 

Although the fully connected autoencoder mimics the PCA model (i.e., with the same matrix multiplications of the PCA algorithm), it shows a lower performance ($\simeq 30\% $ drop of accuracy) than PCA due to two factors. 
First, given the small size of our training set, which negatively affects the data-driven model's performance, it reaches an AUC of only 0.970. Further, the threshold chosen while analyzing only normal data does not permit high accuracy by favoring the specificity. MSE achieved by both anomalies and normal data is very near the test set using frequency domain input data. Therefore, also a small modification in the threshold can impair the accuracy, leading to low sensitivity. 
Note that we choose this threshold with statistical consideration on the validation dataset, ensuring a specificity $>$ 99.9\% on the validation set, but without any assumptions on the sensibility. 
On the other hand, the convolutional autoencoder does not show promising results, with a very low sensitivity of $37.1 \%$.  
This low sensitivity is probably due to the high number of parameters that overfit the training dataset, not allowing it to reach the performance of the other methods.
Comparing frequency and time domains, we first visually analyze the input data.
We notice a slightly different waveform between anomalies and normal data in the time domain, given by higher variations in amplitude and lower frequencies in anomalies.
These changes also noticeable in the power spectrum, with a slight deviation in the first natural component of the viaduct.
Therefore, we feed our algorithm with either raw data or FFT of each input window. Since the viaduct's natural frequency is relatively low, we cut the frequency spectrum between 0-25 Hz.
Although with the FFT pre-processing, we can reach high AUCs of 0.925 and 0.889 for our best models, we see an improvement using time-domain data. Moreover, our unsupervised threshold training does not allow us to reach a satisfactory accuracy on frequency data.
Even though FFT shows a slight difference, it is prone to spectrum leakage due to the measured signal's non-stationary or non-linearity~\cite{kuwalek2020influence}.
To avoid possible spectrum loss, we also evaluate Discrete Wavelet Transform (DWT) approximation coefficients to represent different time and frequency resolutions simultaneously. Thus, we use the DWT coefficients of each 5s window as one other possible input to our anomaly detectors.
DWT results reveal that we can reach as high AUC as FFT with 0.922 and 0.939 for the superior models in the pipeline. Similarly to FFT, due to unsupervised threshold training, DWT does not reach an adequate accuracy, with only $84\%$ and $69.99\%$ for the former algorithms
Table \ref{tab:accuracies} and Fig. \ref{fig:roc} summarize the time (Raw data), frequency (FFT), and time-frequency (DWT) results.
At the end of this exploration, we select the PCA and the time domain as best competitors, and we, therefore, use them in subsequent analysis and in deployment on edge nodes.

\subsection{Hyperparameters exploration}
\subsubsection{Energy Filtering}
\label{subsec:energy_filtering}
\begin{figure}
  \centering
\includegraphics[width=1\columnwidth]{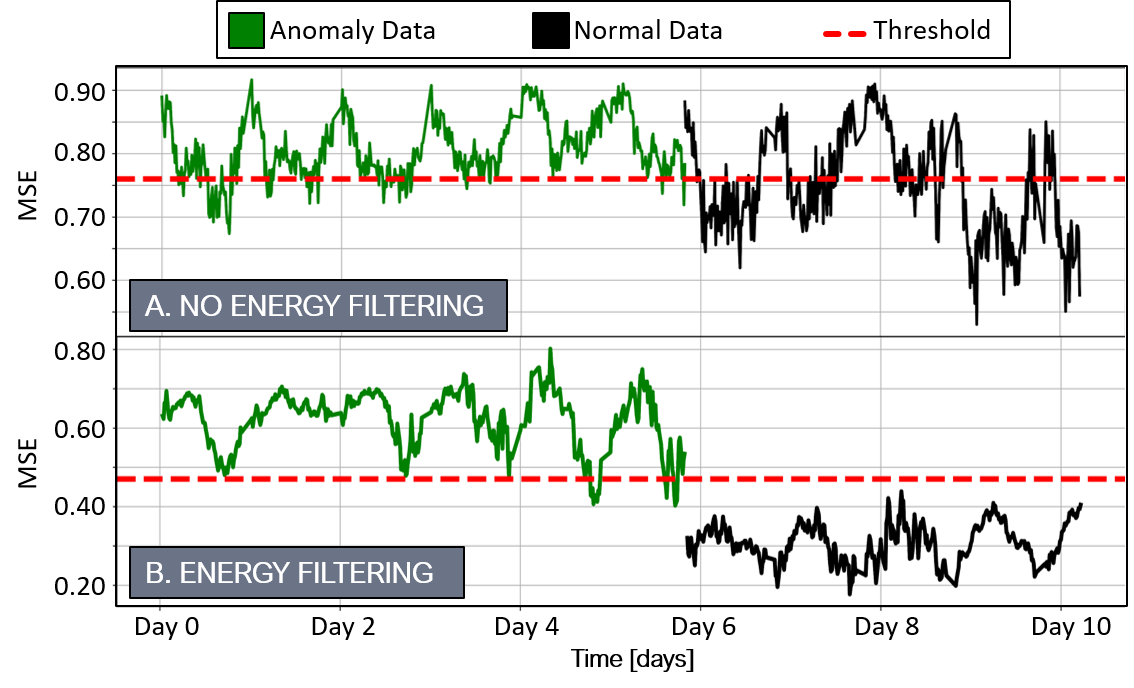}
  \caption{The energy filtering step impacts on the PCA output MSE. In the top panel, we show the MSE when the energy filtering is not applied. In the bottom panel, we show the improved result with its application.}
\label{fig:eg_filtering}
 \vspace{-0.5cm}
\end{figure}
In Sec \ref{sec:scenarios}, we propose filtering non-informative windows to train models with only the most energetic windows, thus removing windows where the viaduct does not vibrate under the passage of vehicles. 
Our hypothesis is that including all the windows leads to higher reconstruction errors of normal and abnormal data, while the gap between the two errors is reduced.
Fig \ref{fig:eg_filtering} quantifies this claim, showing classification with and without the energy filtering block. 
We can observe that the PCA is strongly affected if we omit this filtering step, with a severe drop of specificity/sensitivity (up to $\simeq 41 \% $). 
Notably, the PCA's poor performance is due to the aforementioned increase of MSE of normal windows, whose average move from 0.31 to 0.70.
This experiment confirms our initial idea, given that non-energetic windows only contain white noise, which is not autocorrelated. Thus it is impossible to compress and reconstruct with PCA, leading to high reconstruction errors, similar to anomalies.
Therefore, adding this block allows improving the detector performance strongly.
The energy filtering is not only beneficial for accuracy but also for computation, reducing the total number of processed windows by $\sim17 \%$ on average, thus reducing the total consumed energy.
\subsubsection{Input \& Output Dimension Exploration}
\label{subsec:in_out}
\begin{figure}[t]
  \centering
\includegraphics[width=0.9\columnwidth]{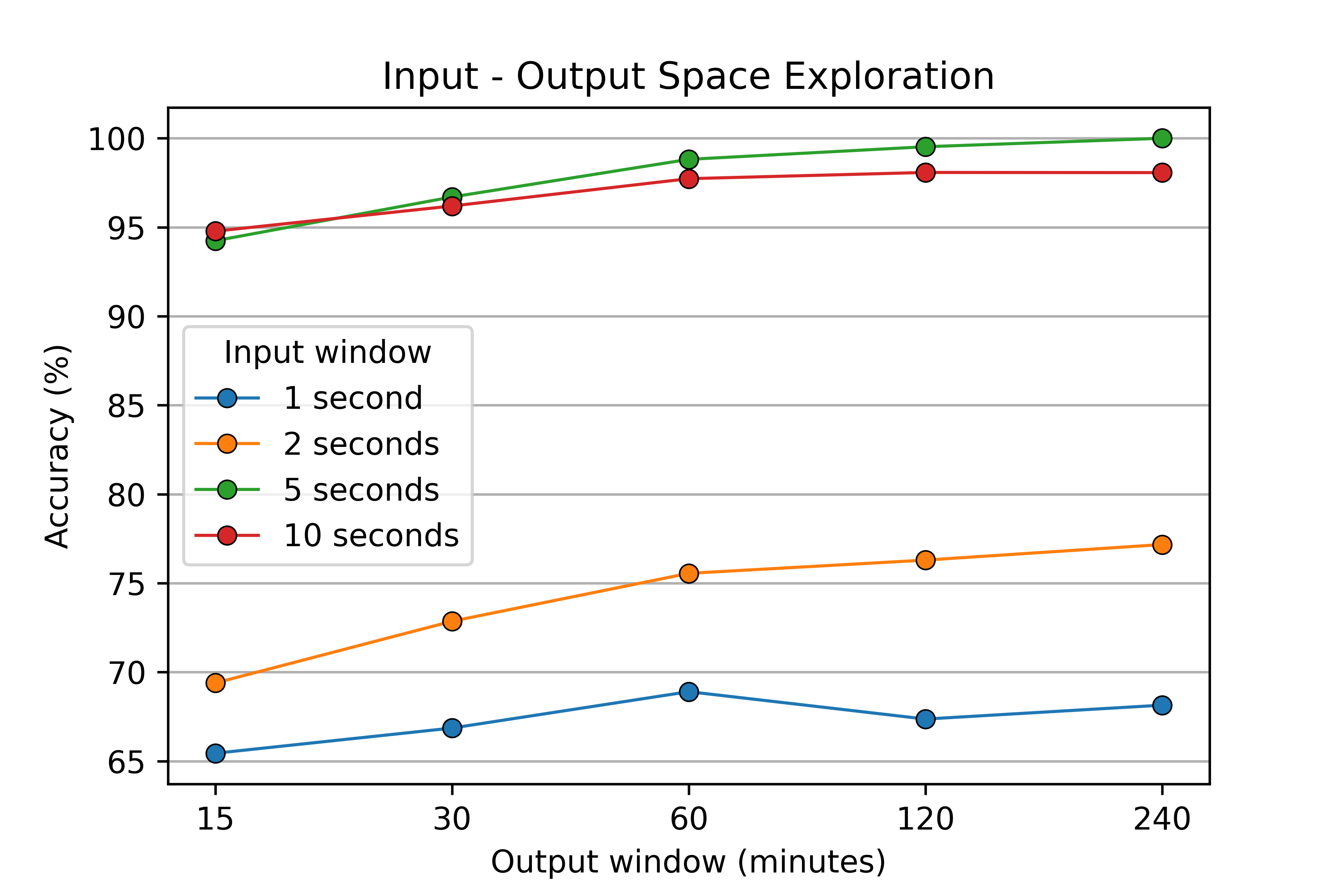}
  \caption{Effect of input-output dimensions sweep on the performance of the best detector (PCA).}
\label{fig:InputOutputExploration}
\end{figure}
Fig. \ref{fig:InputOutputExploration} shows the tuning of input and output dimensions, with twenty different combinations of four input dimensions with five output dimensions.
Input dimension variation is not positively/negatively correlated with algorithm performance.
We can notice that using 5 seconds (grid search between 1,2,5 and 10 seconds) outperforms the other input dimensions values from Fig. \ref{fig:InputOutputExploration}.
On the other hand, using smaller windows reduces the computation and thus the energy consumption of the algorithm execution, leading to a trade-off between energy consumption vs. accuracy.
We will better study this trade-off in the following sections.

Contrary to the input dimension, an increase in the output dimension positively correlates with the framework's performance, resulting in a trade-off in delay vs accuracy. 
However, since viaduct structure modifications are slow, having very low delays is not required.
Therefore, we decided to use 60 minutes of output dimension, which almost saturates performance for 5 seconds windows while having a reasonable delay.
Impressively, increasing the output dimension to 120 and 240 minutes provides further better accuracy. However, the choice of the output dimension, which strongly affects the delay in detecting the status of an anomaly in the viaduct, is related to the specific use-case or necessity of the system. 
For instance, choosing 240 minutes as a dimension leads to the perfect distinction of anomalies and safe time slots (100\% accuracy) but causes a delay of 4 hours in the notification of the damage alarm.
\subsubsection{Compression factor}
\label{subsec:CF}
\begin{figure}
  \centering
\includegraphics[width=0.9\columnwidth]{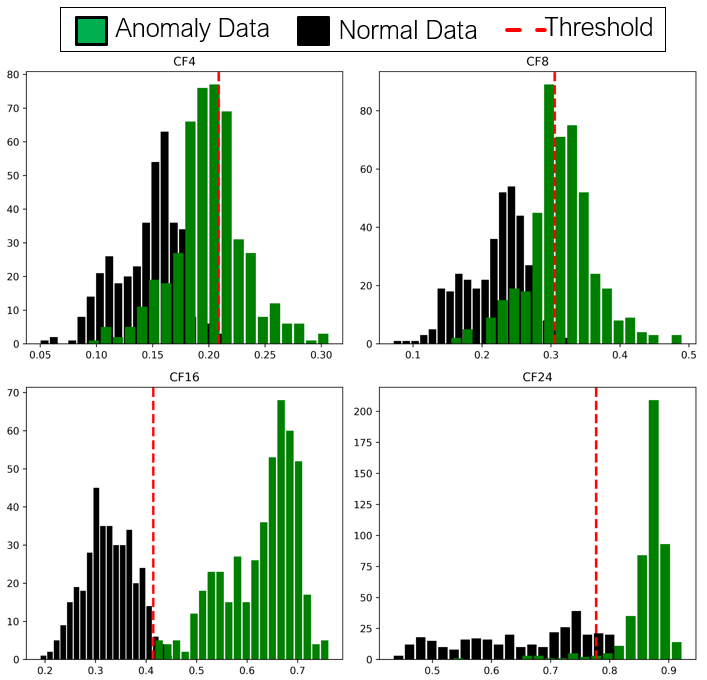}
  \caption{MSE distribution while changing CF parameter}
\label{fig:guass.CF}
 \vspace{-0.5cm}
\end{figure}
We also explore different compression factors for PCA to analyze its effect on framework overall performance. Intuitively, preserving more high-dimensional space elements does not guarantee enhancement in overall performance since they can improve the reconstruction of both normal and abnormal data.           
For this reason, starting from our initial value of 16, we further explore CFs = 4, 8, 24, and 32.

Fig. \ref{fig:guass.CF} shows the distribution of MSE values of anomalies and normal data with four values.
As expected, a lower compression factor leads to an overall better reconstruction of all the data (0.05 - 0.30 MSE with CF = 4), while using a higher CF (CF = 24) causes a higher reconstruction error (0.4-0.9 MSE).
On the other hand, none of these conditions implies higher accuracy, given that the critical metric that leads to high classification accuracy is the gap between the two data distribution.
Exploring the different values, we find that the original CF value, and similar CF = 16, is the sweet spot in this trade-off, leading to a reasonable reconstruction of normal data (0.1-0.4 MSE) and a poor reconstruction of anomalies (0.4-0.8 MSE). 
As can be visually noted, the distance between the means of the two distributions is maximized for CF = 16, with a value of 0.30. Simultaneously, other CFs, 4, 8, and 24, only present 0.05, 0.09, and 0.21 distances, respectively.
Similar to input data dimension, this parameter affects both the computation and memory footprint of the algorithm and will be further explored in the following sections.

\subsection{Synthetic Experiments}
\label{subsec:anomaly_injection}
\begin{figure}
  \centering
\includegraphics[width=0.9\columnwidth]{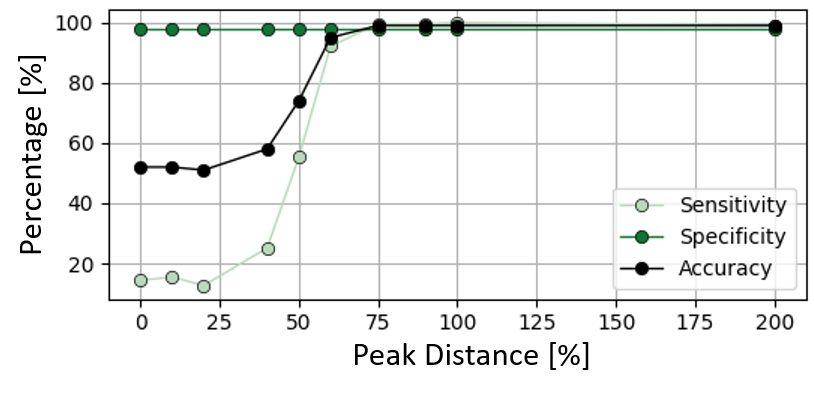}
  \caption{Performance of the PCA classifier while sweeping over several severities of anomalies w.r.t real case scenario of the bridge.}
\label{fig:synthetic}
 \vspace{-0.5cm}
\end{figure}
Lastly, we artificially generated degradations to monitor the robustness of the best-trained detector,i.e., PCA. To inject different sets of anomalies, we modified the distance between the two peaks of normal and anomaly in spectrum density and transformed them back to the time domain. 
We take each 15 minutes of the dataset, process with FFT, and gradually close the two peaks of first natural frequency between anomalies and normal data between 0 to $200 \%$ of the actual distance in the dataset. Note that $100\%$ corresponds to the original real-life anomaly.    
Finally, we transform back the data to the time domain prior and use these new data to test the algorithm.
Reducing the gap between the two peaks allows producing data more similar to normal data, implying a harder task for the detector. 
Fig \ref{fig:synthetic} shows the result of this experiment. Reducing the distance to lower than $75\%$ of the original distance causes the detector to reduce its sensitivity, starting to classify anomalies as normal cases. Note that specificity is constant since the anomaly threshold does not change, given that it is computed only with unmodified normal data. 

\subsection{Model Comparison}
\label{subsec:STOA}

\begin{table}
\centering
\caption{Comparison of our proposed solution with state-of-the-art methods applied to our dataset. The 60 minutes post-processing is identically applied to all methods.}
\label{tab:STOA}

\begin{tabular}{|l|l|l|l|} 
\hline
Method~                                                   & Acc.~                & Spec.                & Sens.      \\ 
\hline
\multicolumn{1}{|l}{\textbf{State-of-the-art algorithms~}~} & \multicolumn{1}{l}{} & \multicolumn{1}{l}{} &            \\ 
\hline
FFT + peaks detection \cite{ref2-42}~~                              & 67.79 \%~            & 99.2\%~              & 43.09 \%   \\
MGD \cite{ref2-46}~                                                 & 59.48 \%~            & 95.66\%~~            & 10.25 \%   \\
AR features + MSD \cite{carden2008arma}~                                   & 58.43 \%~            & 85.90\%~             & 36.82\%   \\
AR features + L1~~                                      & 81.11 \%~~           & 88.49\%~             & 71.80\%    \\ 
\hline
\multicolumn{1}{|l}{\textbf{Our Work}}       & \multicolumn{1}{l}{} & \multicolumn{1}{l}{} &            \\ 
\hline
Raw + PCA                                                       & 98.80\%              & 100\%                     & 97.33\%           \\
DWT + FC Autoencoders~~                                         & 69.99\%~~            & 97.87 \%~            & 47.66\%    \\
FFT + 1D-CNN Autoencoders~~                                     & 56.30\%~~             & 85.28\%~~             & 32.66\%   \\
\hline
\end{tabular}
\end{table}

In this section, we compare our anomaly detectors with methods presented in Sec \ref{sec:relatedWork}.
To do this, we reproduce the pipeline shown in Fig.\ref{fig:pipeline}, substituting the anomaly detection algorithm with the state-of-the-art ones but keeping unchanged the pre-processing and post-processing steps.
We compare the PCA with four other statistical-based approaches based on frequency peak detection \cite{ref2-42}, Multivariate Gaussian Distribution \cite{ref2-46}, and AutoRegressive models \cite{carden2008arma}.
We do not add to the comparison any supervised deep learning methods since they require labels for normal and anomaly cases which are not available at training time in normal SHM use-cases.
Table \ref{tab:STOA} showcases the comparison in terms of accuracy, specificity, and sensitivity.

The literature about anomaly detection in SHM shows that autoregressive moving average (ARMA) residuals are damage-sensitive features of  structures~\cite{carden2008arma,gul2011damage}. Entezami et al. \cite{ref2-18} propose a novel approach to extract these features for big data (GBs). We reproduced their approach on our data, training two different statistical distances, L1 distance and Mahalanobis Square Distance (MSD), to distinguish the normal and anomalous data. Table \ref{tab:STOA} shows that L1 achieves an overall better accuracy (+22.78\%) than Mahalanobis Square Distance (MSD). 
Santos et al. \cite{ref2-42} propose to extract frequency information from the signal and perform the classification based on the position of the main peak of the spectrum. However, in our use case, this method achieves an accuracy of only 67.79\%. This result further proves that frequency features are not suitable to distinguish safe and anomalous time windows on our dataset.
Finally, we investigated a recent study that targets edge computing \cite{ref2-46}, exploiting seven statistical features, (i.e., mean, mean square, variance, standard deviation skewness, kurtosis, and crest factor) together with a multivariate Gaussian model to predict anomalies in vibrating systems. However, also this method fails in our use-case, with a drop in accuracy to $\approx 60\%$. 

In a nutshell, the results in Table \ref{tab:STOA} show that correlation and autocorrelation of 1-D vibrations are promising solutions (exploited by both PCA and AR models) to detect anomalies in viaducts.

\section{Experimental Results: pipeline deployment} 
\label{sec:deployment_results}
\begin{figure*}
  \centering
\includegraphics[width=0.99\textwidth]{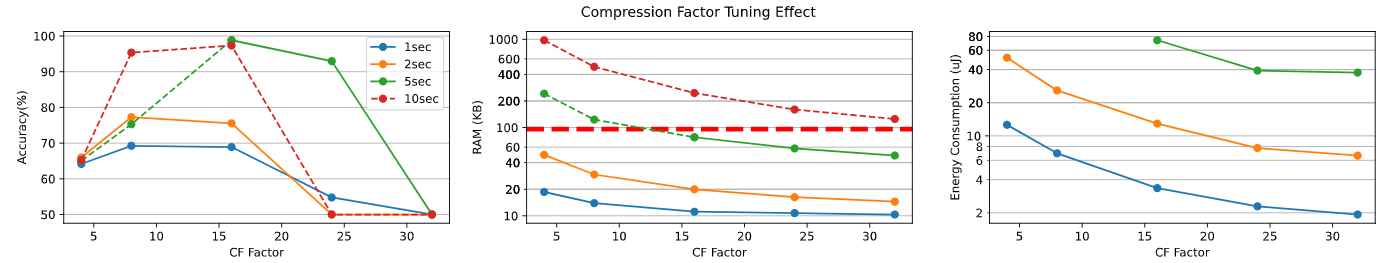}
  \caption{CF tuning versus accuracy, memory, and energy. Horizontal red line points to the limit of MCU memory. Dotted lines represent not deployable solutions.}
\label{fig:CF_deploy}
% \vspace{-0.5cm}
\end{figure*}

This section will analyze several deployments of the best algorithm found in the performance analysis, namely, the PCA, on the processing unit introduced in Sec. \ref{subsec:shmframework}, the STM32L476VGTx. 
All pipeline steps, including data processing, signal reconstruction, and anomaly detection, have been deployed using optimized C code and FreeRTOS operating system.
Initially, we tuned the CF of PCA with multiple input dimensions against memory constraints to realize the utmost limit of PCA deployment with a floating-point compression matrix. We further address each case's energy consumption and execution time to report each case's pros and cons.  
We then fix CF for the best performance and perform interference with MCU to compare its performance with the offline version. 
Finally, we present a comparison of the best solutions to show the pros and cons of the three scenarios discussed in Sec \ref{sec:scenarios}. 
%%%%%%%%%%%%%%%%%%%%%%%%%%%%%%%%%%%%%%%%%%%%%%%%%%%%%%%%%%%%%%%%%%%%%%%%%

%
%%%%%%%%%%%%%%%%%%%%%%%%%%%%%%%%%%%%%%%%%%%%%%%%%%%%%%%%%%%%%%%%%%%%%%%%%%
\subsection{CF tuning Vs. Metric Figures}
\label{subsec:cftuning}
Starting from the accuracy results shown in Sec. \ref{sec:software_results}, we extensively explore the trade-off between accuracy, memory, and energy consumption by modifying both the CF and the input dimension, with a fixed output dimension of 60 minutes. We show the results of our exploration in Fig.\ref{fig:CF_deploy}.
As previously mentioned, CF = 16 results in the best accuracy, with $67.34 \%$, $76.29 \%$ , $98.82 \%$ and $ 97.33 \%$ for input dimensions of 1, 2 ,5 and 10 seconds, respectively.   
Despite the higher accuracy of CF = 16, increasing the CF allows reducing both the memory footprint and energy consumption.
For instance, from the first graph of Fig.\ref{fig:CF_deploy}, we can notice that using CF = 24 with a window of 5 seconds still allows us to reach an acceptable accuracy of $92.97 \%$.
Contrary, using a lower CF causes i) higher energy, ii) higher memory consumption, and iii) lower accuracy, totally excluding these CF values from the trade-off choice.
Therefore, We fix the search space to CF $\in [16,32]$. 
We also remove the 10 seconds input dimension since its compression matrix does not fit the small 96 kB RAM (dotted line in the second graph of Fig.\ref{fig:CF_deploy}).
In this region, we found that the only points that reach an accuracy $ > 80\%$ are achieved for CF = 16 or CF = 24 and input dimension = 5 seconds.
Target application and deployment scenarios can choose the best trade-off between former parameters. Given our pipeline, we found that the largest model that fits the MCU memory is the one with CF = 16 and an input dimension of 5 seconds, achieving  $98.82\%$  accuracy with a  $73.96$uJ energy consumption per inference.

Table \ref{tab:node_deployment} underlines memory footprint, latency, and energy consumption with a fixed CF of 16 and different input dimensions.
Since increasing input dimension corresponds to a more extensive PCA compression matrix, a higher input dimension requires more FLASH space (e.g., 91.04 kB for 5 seconds).
Although the compression matrix is not a problem (roughly 10\% of total FLASH for 5 seconds), the reconstruction procedures occupy up to 77.55 kB (81 \%) of RAM for 5 seconds. Such a high usage area puts a solid constraint for embedding the PCA for larger input dimensions, given the option to run other tasks for data gathering. 
Despite the optimal solution obtained with 5 seconds, reducing the input dimension to 1 second allows us to maintain accuracy of $67.34 \%$, with a reduction of latency of 8.5$\times$ and 9.5$\times$ lower energy consumption.
While the former factor brings no obstacle to the system due to the sampling rate (100 Hz), the latter causes a shorter node lifetime, creating a trade-off between accuracy vs. energy consumption. 
%%%%%%%%%%%%%%%%%%%%%%%%%%%%%%%%%%%%%%%%%%%%%%%%%%%%%%%%%%%%%%%%%%%%%%%%%%%%%
\subsection{Cloud vs. Node costs} 
\label{subsec:final_costs}

\begin{table}
\centering
\caption{Deployment metrics of PCA algorithm with CF = 16, output dimension = 60 minutes, and variable input dimension. Time and Energy are only for one inference.}
\label{tab:node_deployment}
\begin{tabular}{|l|l|l|l|l|l|l}
\hline
Input dim. & FLASH [kB] & RAM [kB] & Time [ms] & Energy [uJ] \\\hline
1                 & 32.82            & 11.12     & 0.754  & 3.35  \\
2                 & 40.63             & 19.95     & 1.568  & 12.9295  \\
\textbf{5}       & \textbf{91.04}    & \textbf{77.55}     & \textbf{6.428}    & \textbf{73.96}   \\
10                & 276.54           & Overflow & -  & -  \\
\hline
\end{tabular}
\end{table}%
\begin{comment}
\begin{table*}
\centering
\caption{NB-IoT deployment cost for the three scenarios of our pipeline\\ 
$E_{sleep} = 390 mJ$ for the three cases. Since our solution energy consumption depends on the vehicle passage, we indicate energy consumption at node with dependency on number of cars passing through the sensor.}
\begin{tabular}{|l|l|l|l|l|}
\hline
Streaming & Network Traffic (KB/H) & Energy Consumption (J)(1 session) & Energy Consumption (J) (1 day) & Energy Consumption (uJ)(at node) \\ \hline
Full              & 780 (600 packets)              & 270.82                         & 6499.83 + E\_sleep            & None                         \\ \hline
Compressed        & 52 (40 packets)                & 18.82                          & 451.68 + E\_sleep             & 21.87 * \#vehicle\_passage     \\ \hline
Prediction        & 0.008 (1 packet)              & 0.450                          & 41.44 + E\_sleep              & 85.9636* \#vehicle passage \\ \hline
\end{tabular}
\end{table*}

\end{comment}

\begin{table*}
\centering
\caption{NB-IoT deployment cost for the scenarios of our pipeline\\ 
$E_{sleep} = 390$ (mJ) is the energy consumption of the node in PSM.\\ $E_{acq} = 52.596$ (mJ) is the energy consumption to acquire 1 second of data}
\begin{tabular}{|l|l|l|l|l|}
\hline
Scenario & Network Traffic (B/H) & NB-IoT E. [J] (1h) & Node Comp. E. [J] (1h) & Gathering E. [J] (1h) \\ \hline
\multicolumn{5}{l}{\textbf{Inference}}\\\hline
Cloud Computation                  &    780 kB    &  248.85 + $E_{sleep}$ &  1.208    & 62.4    \\ \hline
Sens. Inference + Cloud Train       &    3 B   &   0.7130 + $E_{sleep}$ &  0.005    & 62.4      \\ \hline
Sensor Computation                  &    \textbf{3B}    & \textbf{0.7130 + $E_{sleep}$}  & \textbf{0.005} & \textbf{62.4}     \\ \hline
\multicolumn{5}{l}{\textbf{Train}}\\\hline
Cloud Computation                                       &  780 kB   &  248.85 + $E_{sleep}$ &  1.208        & 62.4    \\ \hline
Sens. Inference + Cloud Train                           &  780 kB      &  248.85 + $E_{sleep}$ &  1.208       & 62.4   \\ \hline
Sensor Computation                                      &  \textbf{0 B}        & \textbf{$E_{sleep}$} &  \textbf{0.00162}    & \textbf{62.4}      \\ \hline
\end{tabular}
\label{tab:NBIoT}
\end{table*}

%
%Narrowband IoT (NB-IoT) is a new protocol standardized by 3GPP for Low Power Wide Area, an extension of LTE (4G Long Term Evolution) designed for long battery and low-cost applications where it can virtually work everywhere \cite{Ref3-8}. 

%Figure 6 shows three different possible deployments of our pipeline for training, detection, and updating the training model. We want to quantify gains and losses in each of these scenarios regarding energy consumption and transition costs.
%
%Table 5 indicates streaming data through an NB-IoT module to the cloud. 
%Accelerometer sampling with a 100 Hz sampling frequency rate results in 780 KB samples per hour. Streaming it all to the cloud ends in a lifetime of less than a year since table 5 shows that energy consumption for the transition is high. On the other hand, if we bring all the computational tasks of the detection to the node, only 8 B is necessary to report the structure's status.
%
%\cite{Ref3-10} estimates energy consumption of a node using NB-IoT with terms like transmitted packets and initial connection and disconnecting the module. Then, the data stream to the cloud on periodic sessions (every hour in our case) 
%
%Table 5 shows that streaming all the data to the cloud consumes over 270 J for each session. Our solution reduces this quantity to (mJ) order for a session. On the other hand, putting all the burden on the node introduces power consumption for each MSE extraction window.    
%%%%%%

Narrowband IoT (NB-IoT) is a recent protocol standardized by 3GPP for Low Power Wide Area, an extension of LTE (4G Long Term Evolution) designed for long battery and low-cost applications where it can virtually work everywhere \cite{Ref3-8}. 

NB-IoT consumes more energy per payload packet than other similar technologies. However, since it has no limitation on the number of bytes sent in a single connection to the cell, it is a prominent transmission protocol in the LPWAN category \cite{Ref3-8}. 
NB-IoT deployment of nationally-licensed connectivity (e.g., LTE bands) implies no band usage limitation and no latency for streaming acquired data to the cloud. 
Since in the SHM field, the streaming of data need not to be continuous, and data can be grouped in big batches and sent with a single connection to the cloud, NB-IoT can be an ideal communication option for SHM systems.
Therefore, we use NB-IoT to gauge the benefits of the different scenarios introduced in \cite{Ref3-10}. 

Fig. \ref{fig:scenarios} shows all the options for training and inference. The cloud-based method continuously streamlines the data to the cloud for both the training and detection phases. In contrast, sensor computation only reports the structure's status to the cloud on an hourly basis. We want to quantify each of these scenarios regarding energy consumption and transition costs to develop a scalable solution for SHM applications. 
Tab. \ref{tab:node_deployment} reports the deployment results of model inference at the node. 
The best solution in terms of accuracy, i.e., the PCA with 5 seconds input window dimension, consumes 73.96 uJ. Exploiting a smaller input window dimension, i.e., 1 second, only consumes 3.35 uJ. Although smaller input windows are $3\times$ more energy efficient, the degradation in terms of accuracy compared to bigger ones is too critical. Furthermore, compared to the cloud paradigm analysis presented in table \ref{tab:NBIoT}, the energy consumption of the processing unit is negligible. With this in mind, energy consumption is the only counter effect of larger input dimensions, while other factors like memory footprint and execution time are satisfied. Hence, we keep 5 seconds input dimensions to preserve the performance. 

The node installed on the viaduct works with an output sampling rate of 100 Hz; thus, it generates 100 16-bits samples per second. Therefore, the node generates 200 Bytes per second, leading to 720KB per hour. To estimate this node's energy consumption with the NB-IoT protocol, we use the estimations provided in \cite{Ref3-10}, where diversity in the payload for each packet affects the power consumption of the node. We decided to use a payload of 1300 B for this experiment. This selection of payload can send 650 (1300/2) samples per packet. Hence, we need 554 packets to transmit 720Kb of hourly data. Notice that we send one hour of acquired data all at the same time to leave NB-IoT in the power sleep mode (PSM) for most of the time, reducing the total power consumption.  
However, storing an hour of data further adds a cost of storage ($\approx 1 J/h$) to an off-chip memory (e.g., a micro SD card).
Table \ref{tab:NBIoT} summarizes the energy consumption regarding different sections of both the training and inference part. It shows that exploiting the full deployment of the cloud computing approach consumes 312.848 J/h where it reduces to 63.508 J/h for the localized sensor deployment of our pipeline. An approximate $5\times$ drop of energy consumption for the latter case is due to the low traffic load transmitted to the cloud.               

On the other hand, if we bring all the computation to the node where the node only sends structure's status to the cloud, we can reduce the traffic of the system to only 3B (i.e., "OK" or "NOK"). Given the small number of generated data by the node, for this case, we can tune the payload of the NB-IoT module to only 10 Bytes (smallest possible number) for each packet. Then, we only transmit one packet to the cloud leading to less than 1 J energy consumption. Although our solution reduces transmission cost to the cloud, allowing scalable solutions for large-scale structures, it consumes a high energy rate at the node.  
Table \ref{tab:NBIoT} also reports different sides of transmission vs. node energy consumption trade-off for both training and inference phases. The high traffic rate during inference ($\sim$ 780 kB/hour) for a complete cloud-based approach prohibits its utilization for large-scale SHM scenarios. Contrary, our solution reduces the traffic of only 10B/h to the cloud is extendable to large-scale systems.
On the other hand the node computation energy is always negligible compared to the energy required to gather the acceleration data, thanks to i) the initial energy-filtering of the windows and ii) the lightweight algorithm employed (PCA).
\section{Conclusion} 
\label{sec:Conclusion}
This work proposes an efficient damage detection solution at the edge, simultaneously reducing network traffic and energy consumption, while anomaly detection accuracy is not adversely altered compared to cloud-based systems. First, we propose a new damage detection pipeline, comprising a pre-processing step, an anomaly detection algorithm, and a postprocessing step. Comparing PCA, and two different autoencoders, we show that PCA outperforms the other two methods by approximately $30\%$ and $48\%$, on our SHM dataset collected on a real-standing Italian bridge. We show that by tuning hyperparameters of our pipeline, we further improve the accuracy in the detection of anomalies by $20\%$.

Additionally, we demonstrate the embedding of our tuned pipeline on a tiny low power device, moving the damage detection to the edge of the network. By doing so, we reduce the data traffic by a factor of $\approx8\cdot10^5\times$, from 780 KBytes/hour to 10 Bytes/hour, compared to a cloud-based anomaly detection solution. Further, we reduce the power computation for the node by $5\times$. 

%\section*{Acknowledgment}
%Anonymous

%IEEEtran
\bibliographystyle{IEEEtran}
\bibliography{bibliography.bib}

\section*{Authors}
\begin{IEEEbiography}[{\includegraphics[width=1in,height=1.25in,clip,keepaspectratio]{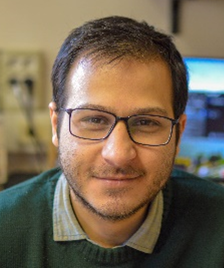}}]{Amirhossein Moallemi}
received his B.Sc degree in Electrical Engineering - Electronics at Zanjan University, Iran, in 2017 and M.Sc degree (cum laude) in Electronic Engineering at the University of Bologna, Italy, in 2020. He is currently working on his Ph.D. degree at the Department of Electrical, Electronic and Information Technologies Engineering(DEI) of the University of Bologna, Italy. His research interests include IoT, low-power hardware and firmware design for embedded systems, machine and deep learning models, Structural Health Monitoring systems.      
\end{IEEEbiography}
\begin{IEEEbiography}[{\includegraphics[width=1in,height=1.25in,clip,keepaspectratio]{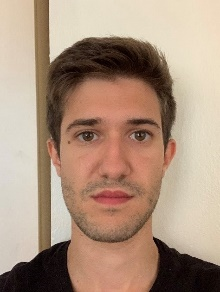}}]{Alessio Burrello}
received his B.Sc and M.Sc degree cum laude in Electronic Engineering and Embedded Systems at the Politecnico of Turin, Italy, in 2016 and 2018.  He is currently working toward his Ph.D. degree at the Department of Electrical, Electronic and Information Technologies Engineering (DEI) of the University of Bologna, Bologna, Italy. His research interests include parallel programming models for embedded systems, machine and deep learning models, hardware oriented deep learning, and code optimization for multicore systems.
\end{IEEEbiography}
\begin{IEEEbiography}[{\includegraphics[width=1in,height=1.25in,clip,keepaspectratio]{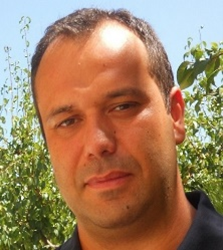}}]{Davide Brunelli}
received his M.S. (cum laude) and Ph.D. degrees in electrical engineering from the University of Bologna, Italy, in 2002 and 2007, respectively.
He is currently an associate professor at the University of Trento, Italy. 
His research interests include IoT and distributed lightweight unmanned aerial vehicles UAV, the development of new techniques of energy scavenging for low-power embedded systems and energy-neutral wearable devices. 
He was leading industrial cooperation activities with Telecom Italia, ENI, and STMicroelectronics.
He has published more than 200 papers in international journals or proceedings of international conferences. He is an ACM member and a senior IEEE member.
\end{IEEEbiography}
\begin{IEEEbiography}[{\includegraphics[width=1in,height=1.25in,clip,keepaspectratio]{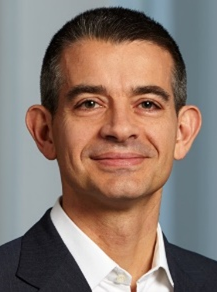}}]{Luca Benini}
holds the chair of digital Circuits and systems at ETHZ and is Full Professor at the Universita di Bologna.
He received a PhD from Stanford University.  He served as chief architect in STmicroelectronics France.
Dr. Benini's research interests are in energy-efficient parallel computing systems, smart sensing micro-systems and machine learning hardware. 
He has published more than 1000 peer-reviewed papers and five books.
He is a fellow of the ACM and a member of the Academia Europaea.
\end{IEEEbiography}

\end{document}